\documentclass[aps,twocolumn,showpacs,fleqn]{revtex4}

\usepackage{graphicx}
\usepackage{amsmath}
\usepackage{amsfonts}
\usepackage{amssymb}

\newcommand{\ket}[1]{|#1\rangle}

\newcommand{\mean}[1]{\langle #1 \rangle}

\newcommand{\Tr}{\text{Tr}\,}
\renewcommand{\Im}{\text{Im}\,}
\renewcommand{\Re}{\text{Re}\,}
\renewcommand{\imath}{\mathrm{i}}

\newcommand{\boldgreek}[1]{\ensuremath{\mbox{\boldmath$#1$}}}
\newcommand{\bdot}{\ensuremath{\,\bullet\,}}

\newcommand{\DM}{\ensuremath{\varrho}}			
\newcommand{\DMR}{\ensuremath{\rho_B}}			
\newcommand{\RDM}{\ensuremath{\sigma}}			

\newcommand{\AuxPsi}{\ensuremath{\tilde{\Pi}}}

\newcommand{\SO}[1]{\ensuremath{\mathcal{#1}}}	

\newcommand{\COMM}[2]{\left[#1, #2 \right]_-}	
\newcommand{\ACOMM}[2]{\left[#1, #2\right]_+}	

\newcommand{\ADO}{auxiliary density operator}

\begin{document}
	\title{Propagation of Time-Nonlocal Quantum Master Equations\\
		     for Time-Dependent Electron Transport}
	\author{Alexander Croy}
	\email{croy@pks.mpg.de}

	\author{Ulf Saalmann}
	
	\affiliation{Max-Planck-Institute for the Physics of Complex Systems,
           N\"{o}thnitzer Str.~38, 01187 Dresden, Germany}
	
	\date{\today}

\begin{abstract}\noindent
Time-resolved electron transport in nano-devices is described by means of a time-nonlocal quantum master equation for the reduced density operator. 
Our formulation allows for arbitrary time dependences of any device or contact parameter.
The quantum master equation and the related expression for the electron current through the device are derived in fourth order of the coupling to the contacts.
It is shown that a consistent sum up to infinite orders induces level broadening in the device.
To facilitate a numerical propagation of the equations we propose to use auxiliary density operators.
An expansion of the Fermi function in terms of a sum of simple poles leads to a set of equations of motion, which can be solved by standard methods.
We demonstrate the viability of the proposed propagation scheme and consider electron transport through a double quantum dot.
\end{abstract}
	
	
	\maketitle
\section{Introduction}\label{sec:Intro}
\noindent
One of the most successful techniques for investigating artificial nano-structures, such as integrated molecules or gated quantum dots, is the measurement of the electron flow through the structure from a source to a drain contact. 
In this field, there has been an enormous progress  in the experimental realization since the first measurements of electric currents through molecules \cite{rezh+97,smno+02} or quantum dots \cite{koma+97,koau+01} more then a decade ago.
Thereby, it has been successfully demonstrated that current measurements provide information equivalent to the traditional spectroscopy of atoms or molecules in the gas phase \cite{sv04}.
In recent years, the possibility to inject ultra-fast voltage pulses into nano-devices paved the way to perform transport measurements beyond steady-state paradigms.
This approach is particularly promising for two reasons:
Firstly, transient relaxation processes can be measured directly in the time domain \cite{hawi+03,fuha+06,pejo+05}.
Secondly, the external control of the pulses can be used to manipulate the electronic dynamics in quantum-dot systems \cite{hafu+03,pejo+05,kobu+06,shha+09,pepe+10,hama+10}, which may eventually be used to realize quantum gates \cite{lodi98}.
In either case an explicit time-dependent theoretical description of the ultra-fast dynamical response
is mandatory for interpreting time-resolved data. Further, it would possibly allow to develop new manipulation schemes for nanoscale devices.

Addressing these challenges there has been considerable progress towards having
real-time methods for the description of electron transport.
Many of these are based on non-equilibrium Green functions (NEGF)
\cite{jawi+94,kust+05,zhma+05,hohe+06,mogu+07,stpe+08,myst+09,crsa09a}.
Hereby, electron-electron interaction is either fully neglected or approximately taken into account by means of self-energies obtained from many-body perturbation theory \cite{myst+09}. 
An alternative approach for strongly interacting systems is given by considering the time evolution of the (reduced) many-particle density matrix \cite{brsc94,leko+02,ovne04,pewa05,wesc+06,jizh+08,moma+09,crsa10a}.
As in the theory of open quantum systems \cite{brpe02}, one studies the reduced density matrix describing exclusively the degrees of freedom of a particular sub-system (here the nano-device) which is embedded in a larger system containing an environment (here the contacts).
Instead of the bosonic environment of a system coupled to a heat bath, the nano-device is coupled to thermal fermionic reservoirs through source and drain contacts. The tunnel coupling of these contacts to the system leads to an exchange of electrons with the reservoirs, and thus to an electric current through the device.
This current, as well as related quantities like level occupations, can be determined by means of the reduced density matrix \cite{wesc+06}.
Its time evolution is given by so-called (generalized) quantum master equations (QME), in which typically the coupling to the environment is incorporated in a perturbative description, often in second order.
 
In this article, we present a propagation scheme which allows for an efficient real-time description of nano-systems with arbitrary time-dependences of any parameter of the system and the contacts, such as level energies, tunnel couplings, or voltages.
We discuss a \emph{fourth-order method}, which serves two purposes:
Firstly, it shows corrections beyond the ``standard description'' with second-order schemes \cite{leko+02,wesc+06,moma+09}.
Secondly, it devices a method for summing up certain terms up to infinitely many orders.
Thereby, the broadening of system levels due to the coupling to the contacts is properly taken into account, which is shown to be crucial for obtaining meaningful results
from solving QMEs in the time domain.

Additionally, we propose the application of \emph{auxiliary density operators\/} by which the calculation of the time evolution is reduced to the propagation of coupled differential equations containing only quantities with one time argument \cite{kl04,wesc+06}.
By means of a recently proposed partial fraction decomposition of the Fermi function \cite{crsa09crsa10}
the number of equations to be propagated is considerably smaller than in previous implementations of the auxiliary-operator method \cite{wesc+06,jizh+08}.

In the following we define the setup of system, contacts and their coupling by giving the
corresponding Hamilton operators. 
In Sect.~\ref{sec:DensityMatrix} we summarize the derivation of the QME for the reduced density matrix up to fourth order. We show how the current can be calculated in terms of reduced (i.\,e.\ system) quantities. 
Auxiliary operators are introduced in Sect.~\ref{sec:AuxOpMeth}. We derive their equations of motion and give an expression for calculating of the electric current.
In Sect.~\ref{sec:InfiniteDensityMatrix} we discuss the extrapolation to infinite orders.
Finally, we present in Sect.~\ref{sec:Appl} a numerical application of our propagation scheme to modeling double quantum dots.

\subsection{Setup}
We consider the usual threefold setup for the
description of tunneling problems, i.\,e.\ a device which is
coupled to two electron reservoirs via tunneling barriers
\cite{caco+71,wija+93}. Accordingly, the total Hamiltonian can be split into three
parts, 
\begin{equation}
	H = H_\mathrm{S} + H_\mathrm{R} + H_\mathrm{SR}\,.
	\label{eq:totalHam}
\end{equation}
The device Hamiltonian 
\begin{equation}\label{eq:DevHamilOp}
  H_\mathrm{S} = \sum_l \varepsilon_l(t) c^\dagger_l c_l
  +\sum_{l\ne m} V_{lm}(t) c^\dagger_l c_m 
  + H_{\rm I},
\end{equation}
is characterized by discrete levels $\varepsilon_l$ which may be
coupled through $V_{lm}$ as, e.g., levels from different parts
in a quantum dot system. In Eq.\ \eqref{eq:DevHamilOp} and below the operators
$c^\dagger_l$ and $c_l$ denote the creation and annihilation of an electron in state $l$.
Electron-electron interactions in the device are accounted for by the interaction
Hamiltonian $H_{\rm I}$. Its particular form depends on the application; for example
a common choice in the context of quantum dots is given by
\begin{equation}
  H_{\rm I} = \frac{1}{2} \sum\limits_{m\ne l} U_{m l} c^\dagger_m c_m c^\dagger_l c_l \;.
\end{equation}
Here, the matrix $U_{m l}$ accounts for intra and inter-dot interactions, if
the labels $m$ and $l$ belong to the same or different quantum dots, respectively.
The reservoir Hamiltonian
\begin{equation}
  H_\mathrm{R} = \sum_{\alpha \in \mathrm{L},\mathrm{R}} 
  \sum_k \varepsilon_{\alpha k}(t) b^\dagger_{\alpha k} b_{\alpha k}\;.
  \label{eq:ResHamilOp}
\end{equation}
describes non-interacting electrons in the left
($\alpha=\mathrm{L}$) and right ($\alpha=\mathrm{R}$) contact.
The respective tunneling Hamiltonian is given by
\begin{subequations}\label{eq:TunnHam}\begin{align}
  H_\mathrm{SR}
  &= \sum_l\left\{\sum_\alpha B^\dagger_{\alpha l} c_l
    + \rm{h.c.}\right\}
  \label{eq:TunnHamB}\\
  B^\dagger_{\alpha l} & = \sum_k  T^\alpha_{k l} b^\dagger_{\alpha k}
  \label{eq:Boperators}
\end{align}\end{subequations}%
with $T^\alpha_{k l}$ denoting the couplings between device and
reservoir. 
The operators $b^\dagger_{\alpha k}$ and $b_{\alpha k}$ are
electron creation and annihilation operators for reservoir
states.
The abbreviations \eqref{eq:Boperators} are sums of reservoir operators which will be
convenient later on.  
In general, as emphasized by the time arguments, all levels and
couplings may explicitly depend on time through the variation of source,
drain or gate voltages. 

In order to simplify the notation we will use a compact multi-index notation
in the following. Thereby, the multi-index 
$\mathbf{a} = (a_0,a_1,a_2)$
encloses the operator type, either annihilation or creation
operator, $a_0=+,-$, contact label
$a_1\equiv\alpha=\mathrm{L},\mathrm{R}$, and system state
$a_2\equiv l$.
A sum over the multi-index $\sum_\mathbf{a}$ corresponds to a
threefold sum $\sum_{a_0}\sum_\alpha\sum_l$.
Further, the system operators will be denoted by $S_{\mathbf{a}}$ and the
reservoir operators, as introduced in Eq.~(\ref{eq:Boperators}),
by $B_{\mathbf{a}}$ with 
\begin{equation}\begin{array}{rclcrcl}
  B^{(+)}_{\alpha l} &=& B^\dagger_{\alpha l} && B^{(-)}_{\alpha l} &=& B_{\alpha l} \;,\\
  S^{(+)}_{\alpha l} &=& c_l &\qquad& S^{(-)}_{\alpha l} &=& c^\dagger_l \;.
\end{array}\end{equation}
Therewith, the tunneling Hamiltonian (\ref{eq:TunnHam}) becomes
\begin{equation}
	H_{\rm SR} = \sum\limits_{\mathbf{a}} a_0 B_{\mathbf{a}} S_{\mathbf{a}} \;.
\end{equation}
The component $a_0$ accounts for the fermionic character of the operators.
Throughout the article we set $\hbar = 1$.

\section{Reduced Density Matrix}\label{sec:DensityMatrix}
\noindent
The evolution of the full density matrix $\DM(t)$ is given by 
the Liouville-von Neumann equation, which in
the interaction picture with respect to $H_{\rm S} + H_{\rm R}$
reads \cite{brpe02,maku00}
\begin{equation}\label{eq:FullDensEq}
	\imath \frac{\partial}{\partial t}\DM(t) = \COMM{H_\mathrm{SR}(t)}{\DM(t)} \equiv \SO{L}(t)\DM(t) \;.
\end{equation}
In the rightmost expression we introduced the super-operator
$\SO{L}(t)$ for the commutator with the tunneling
Hamilton operator. 
Back transformation to the Schr\"{o}dinger representation is
achieved by the super-operator
\begin{subequations}
  \label{eq:SupOpU}
\begin{equation}
  \SO{U}_{\rm 0}(t,\tau) \bdot = U_{\rm 0}(t,\tau) \bdot U^\dagger_{\rm 0}(t,\tau)
\end{equation}
with
\begin{equation}
  U_{\rm 0}(t,\tau) ={\cal T}\exp\left(-\imath\int_\tau^t 
    dt'\left[H_\mathrm{S}(t') + H_\mathrm{R}(t')\right]\right)\,.
\end{equation}
\end{subequations}
Considering only the degrees of freedom of the device, it is convenient to
define a {\it reduced density matrix} $\RDM(t) = \Tr_{\rm B}\{ \DM(t) \}$
in the usual way.
Starting from Eq.~(\ref{eq:FullDensEq}) an equation of motion for
the reduced density matrix $\RDM(t)$ might be found in terms of cumulant expansions.
Here we will use the {\it partial cumulants},
which yields a time-nonlocal generalized QME \cite{te74,mo65,na58,zw60}. 
The respective equation of motion is
\begin{align}
	\imath \frac{\partial}{\partial t}\RDM(t) 
		=& -\imath \int\limits_{t_0}^{t} d\tau \; \SO{K}(t,\tau) \RDM(\tau) \notag \\
		=& -\imath \sum\limits_{n=1}^{\infty} (-1)^{n-1} \int\limits_{t_0}^{t} d\tau \; \SO{K}^{(2n)}(t,\tau) \RDM(\tau) 		
	\;.
	\label{eq:TNLQMEIP}
\end{align}
The memory kernel $\SO{K}(t,\tau)$ has been expanded in partial cumulants
with the $2n$-th order memory kernel given by
\begin{align}\label{eq:2nKernelPC}
	\SO{K}^{(2n)}(t,\tau) =& \int\limits_{\tau}^t d\tau_1 \ldots \int\limits_{\tau}^{\tau_{2n-3}} d\tau_{2n-2}  \\
							&\quad\times\mean{\SO{L}(t)\SO{L}(\tau_1) \ldots \SO{L}(\tau_{2n-2})\SO{L}(\tau)}_{\rm pc}\notag
\end{align}
and the partial cumulants $\mean{\ldots}_{\rm pc}$ being defined via a recursion 
relation \cite{te74}. Note, that in Eq.\ \eqref{eq:TNLQMEIP} we have anticipated the vanishing of the odd order cumulants due to our choice of the initial density operator, which will be discussed below.
Obviously, the $2n$-th order memory kernel contains $n$ super-operators $\SO{L}$
and therefore $n$ times the tunnel coupling $H_{\rm SR}$. 
It has been shown \cite{ti08} how this expansion can be connected to a diagrammatic
description \cite{scsc94}, which provides a clearer interpretation in terms of tunneling processes.
However, for practical calculations one usually considers only the second order QME.
This is mainly due to the complicated structure of nested integrals in Eq.\ \eqref{eq:2nKernelPC}. In the following we will provide explicit expressions of the
second and fourth-order kernels and in next section we will show how an {\it auxiliary operator method} can be utilized to deal with the integrals.

\subsection{Second and Fourth Order Memory Kernels}
Firstly, we only consider the second order and fourth-order kernels, 
which explicitly read
\begin{subequations}
\label{eq:mk24}  
\begin{align}
	\SO{K}^{(2)}(t,\tau) &= \mean{\SO{L}(t)\SO{L}(\tau)} \;,\\
	\SO{K}^{(4)}(t,\tau) &= \int\limits_{\tau}^t d\tau_1 \int\limits_{\tau}^{\tau_1} d\tau_2 
								\mean{\SO{L}(t)\SO{L}(\tau_1)\SO{L}(\tau_2)\SO{L}(\tau)} \notag\\
						&\qquad		-\mean{\SO{L}(t)\SO{L}(\tau_1)}\mean{\SO{L}(\tau_2)\SO{L}(\tau)}  \;.
\end{align}
\end{subequations}
Here the averages $\mean{\ldots}$ are meant to be taken with respect
to the equilibrium reservoir state, i.\,e.\ 
$\mean{\ldots} = \Tr_{\rm R} \left\{ \ldots \DMR(t_0) \right\}$.
Notice that in writing Eq.~\eqref{eq:TNLQMEIP} we have assumed that the initial full density matrix factorizes, i.\,e.\
$\DM(t_0) = \DMR(t_0) \RDM(t_0)$, where the reservoirs are taken to be in equilibrium (grand canonical ensemble). Therefore,
taking the initial time to the infinite past, the QME
description corresponds to the scheme introduced by Caroli and co-workers
\cite{caco+71}. 

The second order memory kernel in terms of system and reservoir operators is explicitly given by the following commutator \cite{maku00,brpe02}
\begin{align}\label{eq:2ndkernel}
	\SO{K}^{(2)}(t,\tau) \bdot = \sum\limits_{\mathbf{a}, \mathbf{b}} &
		\COMM{S_{\mathbf{a}}(t)}{ C_{\mathbf{a} \mathbf{b}}(t,\tau) S_{\mathbf{b}}(\tau) \bdot \right. \notag\\
			&\left. -  \bdot S_{\mathbf{b}}(\tau) C_{\mathbf{b} \mathbf{a}}(\tau, t) }\;,
\end{align}
where we used the reservoir correlation functions
\begin{equation}\label{eq:CorrFun}
	C_{\mathbf{a} \mathbf{b}}(t,\tau) = \mean{B_{\mathbf{a}}(t) B_{\mathbf{b}}(\tau)}\;.
\end{equation}
Due to the fermionic character of the $B$-operators, there are
only two possible types of correlation functions
\begin{subequations}\label{eq:TypesCorrFun}\begin{align}
    C_{\alpha m l}^{(+-)}(t,\tau) &= 
    \mean{B^\dagger_{\alpha m}(t) B_{\alpha l}(\tau)} 
    \;, \\
    C_{\alpha l m}^{(-+)}(t,\tau) &= 
    \mean{B_{\alpha l}(t) B^\dagger_{\alpha m}(\tau)}  
    \;,
\end{align}\end{subequations}
which we can identify with the lesser and greater tunneling
self-energies appearing in the NEGF formalism, cf.\ Eqs.~\eqref{eq:defself}.
Note that all correlation functions with operators from
different reservoirs vanish since there is no direct coupling of the
reservoirs.  

For convenience we introduce two {\it correlation super-operators},
\begin{subequations}\label{eq:corrsupop}\begin{align}
    \label{eq:corrsupopC}
    \SO{C}^{(2)}_{\mathbf{a} \mathbf{b}}(t,\tau) \bdot = {} &
    C_{\mathbf{a} \mathbf{b}}(t,\tau) S_{\mathbf{b}}(\tau) \bdot 
    \notag\\ &
    -  \bdot S_{\mathbf{b}}(\tau) C_{\mathbf{b} \mathbf{a}}(\tau, t) \;, \\
    \label{eq:corrsupopA}
    \SO{A}^{(2)}_{\mathbf{a} \mathbf{b}}(t,\tau) \bdot = {} &
    C_{\mathbf{a} \mathbf{b}}(t,\tau) S_{\mathbf{b}}(\tau) \bdot 
    \notag\\ &
    +  \bdot S_{\mathbf{b}}(\tau) C_{\mathbf{b} \mathbf{a}}(\tau, t) \;.
\end{align}\end{subequations}
The first definition allows for writing Eq.~(\ref{eq:2ndkernel}) in
a compact form. According to the two correlation functions
(\ref{eq:TypesCorrFun}), there are also two types of
non-vanishing correlation super-operators (\ref{eq:corrsupopC}). 
The second definition (\ref{eq:corrsupopA}) is introduced here
for completeness.
It will be used for the fourth-order kernel, which we discuss now.

The fourth order is more complicated. We summarize all details in appendix \ref{sec:PartCum}
and give here, by means of Eqs.~\eqref{eq:4thcumulant} and \eqref{eq:4thordered},
only the final expression for the fourth-order kernel
\begin{align}\label{eq:4thkernel}
  \SO{K}^{(4)}(t,\tau) \RDM(\tau) =& \sum\limits_{\mathbf{a}, \mathbf{b}, \mathbf{c}, \mathbf{d}} 
  \COMM{S_{\mathbf{a}}(t)}{ 
    \int\limits_{\tau}^t d\tau_1 \int\limits_{\tau}^{\tau_1} d\tau_2
    \ACOMM{\vphantom{\SO{C}^{(4)}_{\mathbf{a}}}S_{\mathbf{c}}(\tau_1)}{	\right.\right. \notag \\
  &\left.\left.	\SO{C}^{(4)}_{\mathbf{a} \mathbf{c} \mathbf{d} \mathbf{b}}(t,\tau_1,\tau_2,\tau) \RDM(\tau) 
    }\vphantom{ \int\limits_{\tau}^{\tau_1}} 
  }
\end{align}
with a new super-operator
\begin{align}
	\SO{C}^{(4)}_{\mathbf{a} \mathbf{c} \mathbf{d} \mathbf{b}}(t,\tau_1,\tau_2,\tau)  =& 
		\SO{A}^{(2)}_{\mathbf{c} \mathbf{d}}(\tau_1,\tau_2) \SO{C}^{(2)}_{\mathbf{a} \mathbf{b}}(t,\tau)  \notag\\
	&-   \SO{A}^{(2)}_{\mathbf{a} \mathbf{d}}(t,\tau_2) \SO{C}^{(2)}_{\mathbf{c} \mathbf{b}}(\tau_1,\tau)  \;,
 \;,
\end{align}
where we have used the correlation super-operators \eqref{eq:corrsupop}.
It should be noted that each of the sums in \eqref{eq:4thkernel} runs over a three-fold multi-index,
which comprises the operator type, the reservoir and the system state.

\subsection{Electric Current}\label{sec:Current}
In order to get a complete description of the nanodevice dynamics an expression
for the time-dependent electric current is mandatory. In the following we will show
that such an expression may be found in terms of reduced quantities.
The electric current through tunneling barrier $\alpha$ is given by the rate
of change of the particle number in the respective reservoir
\cite{wija+93,brsc94},
\begin{align}
  J_\alpha (t) 
  &= -e \frac{d}{dt}\mean{N_\alpha } 
  = -\imath e \mean{ \COMM{H}{N_\alpha} } \notag\\
  &= -{2 e\;} \Im \left\{ \sum_l \Tr\left(
      B^\dagger_{\alpha l}c_l \DM(t) \right) 
  \right\} \notag\\
  &= {2 e\;} \Re \left\{  \Tr_{\rm S} \Tr_{\rm R} \sum_l \left(
      \imath B^\dagger_{\alpha l}(t) c_l(t) \DM(t) \right) 
  \right\} \;,
\end{align}
with the number operator for reservoir $\alpha$ given by $N_\alpha = \sum\limits_k b^\dagger_{\alpha k} b_{\alpha k}$.
In the last line we have used the invariance of the trace against unitary
transformations to switch to the interaction picture with respect to $H_0$.
From the equation above one sees that the electron current is given by
a block of the total one-electron density matrix describing the coherence
between reservoir and system. The derivation follows the arguments
given for the expansion of the memory kernel in terms of partial cumulants,
which makes use of projection operators 
\footnote{Taking projection operator $\SO{P}\bdot = \DMR \Tr_{\rm R} \bdot$
and its complement $\SO{Q}$, inserting $\mathbf{1} = \SO{P}+\SO{Q}$
between $B^\dagger_{\alpha l}(t) c_l(t)$ and $\DM(t)$, and using the
formal solution for $\SO{Q}\DM(t)$ directly 
leads to the expression given in Eq.~(\ref{eq:CurrPartCum}). See also
Ref.\ \cite{cr10}.
}.
It can be shown that the current is given by the following expression \cite{cr10}
\begin{widetext}
\begin{equation}\label{eq:CurrPartCum}
  J_\alpha (t) = {2 e\;} \Re \left\{ \Tr_{\rm S} \left[
      \sum\limits_{n=1}^{\infty} (-1)^{n-1} \int\limits_{t_0}^{t} d\tau \;
      \int\limits_{\tau}^t d\tau_1 \ldots \int\limits_{\tau}^{\tau_{2n-3}} d\tau_{2n-2} 
      \mean{H^{(+)}_{{\rm SR}, \alpha}(t)\SO{L}(\tau_1) \ldots \SO{L}(\tau_{2n-2})\SO{L}(\tau)}_{\rm pc}\,
		\RDM(\tau)
		\right] \right\} \;,
\end{equation}
\end{widetext}
with $H^{(+)}_{{\rm SR}, \alpha}(t) = \sum_l B^\dagger_{\alpha l}(t) c_l(t)$.
The integrand is very similar to the expanded memory kernel \eqref{eq:2nKernelPC},
which will be used in the next section to calculate the current from auxiliary
density operators.

\section{Auxiliary Operator Method}\label{sec:AuxOpMeth}
\noindent
In the following we device an approach which allows for the efficient propagation of the QME for the reduced density matrix $\RDM$, cf.\ Eq.\,\eqref{eq:TNLQMEIP}. 
We introduce \emph{auxiliary density operators\/} (ADOs) which contain a time integration of the memory kernel operators $\SO{K}^{(2)}$ and $\SO{K}^{(4)}$  applied on $\RDM$. 
\def\ADO{ADO} 
Most importantly, these \ADO{}s depend only on \emph{one\/} time argument which is why they are much easier to handle numerically. Their time evolution is given by one equation of motion.

\subsection{Auxiliary density operators (ADOs)}

Using the $\SO{C}$ super-operators (\ref{eq:corrsupopC}) we
define second order \ADO{}s
\begin{equation}\label{eq:Def2ndAuxOp}
  \Pi^{(2)}_{\mathbf{a}}(t) = \sum\limits_{\mathbf{b}} \int\limits_{t_0}^{t} d\tau \; 
  \SO{U}_{\rm S}(t) \SO{C}^{(2)}_{\mathbf{a} \mathbf{b}}(t,\tau) \RDM(\tau) \;,
\end{equation}
which contain information on the coherence between system and reservoirs
as we will show in Sec.~\ref{sec:CurrentADO}. Notice that the back transformation 
$\SO{U}_{\rm S}$ to the Schr\"{o}dinger representation is also contained
in Eq.~\eqref{eq:Def2ndAuxOp}; the respective super-operator $\SO{U}_{\rm S}$ is defined as in
Eq.~(\ref{eq:SupOpU}) but with $H_{\rm S}$ only.
By means of these ADOs, the second order time-nonlocal QME in Schr\"{o}dinger representation can be compactly written as
\begin{equation}\label{eq:EOMTNLQME}
	\imath \frac{\partial}{\partial t} \RDM(t) =
		\COMM{H_{\rm S}}{\RDM(t)} 
		-\imath \sum\limits_{\mathbf{a}} 
			\COMM{S_{\mathbf{a}}}{ \Pi^{(2)}_{\mathbf{a}}(t)} \;.
\end{equation}
This form is identical to the standard one
\cite{brpe02,maku00}, which becomes clear if one of the \ADO{}s 
is written explicitly, e.\,g.\ $\mathbf{a} = (+,\alpha,m)$
\begin{align}
  \Pi^{(2,+)}_{\alpha m}(t) 
  = {} & \sum\limits_{l} \int\limits_{t_0}^{t} d\tau  \\
  & \left( C^{(+-)}_{\alpha l m}( t, \tau ) \,U_{\rm S}(t,\tau)\, c_l^{\dagger} \RDM(\tau)\, U^\dagger_{\rm S}(t, \tau) \right. \nonumber\\
  & \left. -
    C^{(-+)}_{\alpha m l}( \tau, t ) \, U_{\rm S}(t,\tau)\, \RDM(\tau) c_l^{\dagger}\, U^\dagger_{\rm S}(t, \tau) \right) \;.\nonumber
\end{align}
Before discussing the propagation of the equations for the
reduced density matrix $\sigma(t)$ and the operators
$\Pi^{(2)}_{\mathbf{a}}(t)$ we will derive the equations for the
next higher order.

In analogy to the second order case we introduce a fourth-order \ADO
\begin{align}\label{eq:Def4thAuxOp}
  \Pi^{(4)}_{\mathbf{a}}(t) =& \sum\limits_{\mathbf{b}, \mathbf{c}, \mathbf{d}} 		
  \int\limits_{t_0}^t d\tau \int\limits_{\tau}^t d\tau_1 \int\limits_{\tau}^{\tau_1} d\tau_2 \;
  \SO{U}_{\rm S}(t) \notag\\
  &\times \ACOMM{S_{\mathbf{c}}(\tau_1)}{ 
    \SO{C}^{(4)}_{\mathbf{a} \mathbf{c} \mathbf{d} \mathbf{b}}(t,\tau_1,\tau_2,\tau) \RDM(\tau) 
  } \;,
\end{align}
which allows for a compact form of the fourth order kernel (\ref{eq:4thkernel}).
Thus we can write the QME up to fourth order in Schr\"{o}dinger
representation as
\begin{align}
  \imath \frac{\partial}{\partial t} \RDM(t) = {} &
  \COMM{H_{\rm S}}{\RDM(t)} \nonumber\\ &
  -\imath \sum\limits_{\mathbf{a}} 
  \COMM{S_{\mathbf{a}}}{ \Pi^{(2)}_{\mathbf{a}}(t) - \Pi^{(4)}_{\mathbf{a}}(t)}\;.
  \label{eq:EOMTNL4thQME}
\end{align}
with the $\Pi^{(n)}$ operators defined by Eqs.~\eqref{eq:Def2ndAuxOp} and (\ref{eq:Def4thAuxOp}).
Having this compact form we may now derive equations of motion for the \ADO{}s
in the so-called wide-band limit.

\subsection{Wide-band Limit (WBL)}\label{sec:wbl}
The proposed propagation method is based on the observation that a closed set
of equations of motion for $\RDM$ and $\Pi^{(2)}$ can be obtained, if the 
reservoir correlation functions \eqref{eq:TypesCorrFun} can be expanded in
a sum of exponentials \cite{meta99,wesc+06}. In particular, this is possible
if the so-called level-width function,
\begin{equation}\label{eq:DefLevelWidth}
  \left[\boldgreek{\Gamma}_\alpha (\varepsilon) \right]_{l m}
  = \sum\limits_k T^\alpha_{k l} T^{\alpha *}_{k m} \delta (\varepsilon_{\alpha k} -\varepsilon ) \;,
\end{equation}
can be approximated by a sum of Lorentzians \cite{meta99}. If $\boldgreek{\Gamma}_\alpha$
is only slowly changing with $\varepsilon$ on the energy scale given by the system,
one can employ the wide-band approximation, i.\,e.~assuming an energy independent
level-width function, $\boldgreek{\Gamma}_\alpha (\varepsilon) = \boldgreek{\Gamma}_\alpha$.
For the rest of this article we will concentrate on this limit. 
Using an expansion of the Fermi-Dirac function as described below, one 
still finds the correlation functions to be sums of exponentials plus one term being proportional
to a $\delta$-function in the time domain, which is a direct consequence of the WBL.
To see this explicitly, we consider the calculation of the reservoir correlation functions
(\ref{eq:CorrFun}), which involves an integration over the Fermi
function, see Eqs.~(\ref{eq:TypesCorrFun}),
\begin{align}
  C_{\mathbf{a} \mathbf{b}}(t,\tau) & =
  \left<\mathrm{e}^{+\imath H_\mathrm{R}t}B_\mathbf{a}\mathrm{e}^{-\imath H_\mathrm{R}t}
    \mathrm{e}^{+\imath H_\mathrm{R}\tau}B_\mathbf{b}\mathrm{e}^{-\imath H_\mathrm{R}\tau}\right>
  \nonumber\\
  & =  \int\frac{d\varepsilon}{2 \pi}
  \boldgreek{\Gamma}^{a_0}_\alpha(\varepsilon,t,\tau)
  f(a_0\beta(\varepsilon{-}\mu_\alpha))
  \mathrm{e}^{a_0\imath\varepsilon (t-\tau)}\,.
  \nonumber
\end{align}
The sign in the Fermi function $f$ and in the exponential
characterizes the type of the correlation function, 
namely $a_0=+$ refers to $C^{(+-)}$ and $a_0=-$ to $C^{(-+)}$,
cf.\ Eqs.~(\ref{eq:TypesCorrFun}). As usual $\beta$ denotes the inverse temperature.
The level-width function as defined in Eq.~\eqref{eq:DefLevelWidth} also
depends on the order of the bath operators. We follow the usual convention, i.e.
\begin{equation}
  \Gamma_{\mathbf{a} \mathbf{b}} = \Gamma_{\mathbf{b} \mathbf{a}} = \boldgreek{\Gamma}^{a_0}_\alpha = 
  \begin{cases}
    \boldgreek{\Gamma}_\alpha &\text{for}\quad a_0 = - \\
    \boldgreek{\Gamma}^\dagger_\alpha &\text{for}\quad a_0 = + 
  \end{cases}\;.
\end{equation}
For convenience we also neglect an explicit time-dependence of the tunneling matrix elements.
Then, in the WBL the correlation functions read
\begin{equation}\label{eq:CorrFunInteg}
  C_{\mathbf{a} \mathbf{b}}(t,\tau) 
  = \boldgreek{\Gamma}^{a_0}_\alpha
  \int\frac{d\varepsilon}{2 \pi} f(a_0\beta(\varepsilon{-}\mu_\alpha))
  e^{a_0\imath\varepsilon (t-\tau)}\,.
\end{equation}
In order to perform the energy integration we expand the Fermi function 
in terms of a finite sum over simple poles
\begin{align}
  f(\pm\beta(\varepsilon{-}\mu_\alpha)) & =
  \frac{1}{1+e^{\pm\beta(\varepsilon-\mu)}}
  \label{eq:ExpFermiFun}\\
  & \approx \frac{1}{2}\mp\frac{1}{\beta}\sum_{p=1}^{N_\mathrm{F}}\left(
    \frac{1}{\varepsilon{-}\chi_p^+}
    +\frac{1}{\varepsilon{-}\chi_p^-}\right)
  \nonumber
\end{align}
with $\chi_{\alpha p}^+=\mu_\alpha + x_p/\beta = \left( \chi_{\alpha p}^- \right)^*$.
Instead of using the Matsubara expansion \cite{ma90},
with poles $x_p=\imath\pi(2p{-}1)$,
we use a partial fraction decomposition of the 
Fermi function \cite{crsa09crsa10}, which converges much faster than
the standard Matsubara expansion.
Furthermore it allows to estimate the error made by truncating
the sum (\ref{eq:ExpFermiFun}) at $N_\mathrm{F}$ terms.
For this decomposition the poles $x_p=\pm 2\sqrt{z_p}$ are given by
the eigenvalues $z_p$ of an ${N_\mathrm{F}}{\times}{N_\mathrm{F}}$ matrix \cite{crsa09crsa10}.
We take the branch of the root $\sqrt{z_p}$ such that 
$\Im(x_p)>0$ for all $p$. 
Thus all poles $\chi_p^+$ ($\chi_p^-$) are in the upper (lower)
complex plane.

Finally, one obtains by means of Jordan's Lemma
the following expressions,
\begin{subequations}\label{eq:CorrFunAsSum}\begin{align}
    C_{\mathbf{a} \mathbf{b}}(t,\tau) 
    & =
    \frac{1}{2} {\Gamma}_{\mathbf{a}\mathbf{b}} \delta (t{-}\tau)
    -\sum\limits^{N_\mathrm{F}}_{p=1}
    \mathbf{C}_{\mathbf{a} \mathbf{b} p} (t,\tau) \,,
    \label{eq:CorrFunSum}\\
    \mathbf{C}_{\mathbf{a} \mathbf{b} p} (t,\tau)
    & = \frac{s\,\imath}{\beta} {\Gamma}_{\mathbf{a}\mathbf{b}} 
    \mathrm{e}^{\imath\chi_{\mathbf{a} p}(t{-}\tau)}\,.
     \label{eq:CorrFunTerm}\\
    \mathbf{C}_{\mathbf{a}\mathbf{b} p} (t,t\mp0) &= \frac{\pm\imath}{\beta} 
    {\Gamma}_{\mathbf{a}\mathbf{b}}
\end{align}\end{subequations}
with $s=\mathrm{sgn}(t{-}\tau)$ and $\chi_{\mathbf{a} p}=a_0\chi_{\alpha p}^{s a_0}$.
Apart from the first term in Eq.\ (\ref{eq:CorrFunSum}) the
correlation functions are sums over exponentials.
The same strategy might be used if the level-width function $\boldgreek{\Gamma}_\alpha(\varepsilon)$
is given by a sum of Lorentzians. In this case the first term in Eq.~\eqref{eq:CorrFunSum} is also
a sum of exponentials; see Ref.~\cite{meta99, wesc+06, jizh+08}.

If the energies in the reservoir Hamiltonian \eqref{eq:ResHamilOp} are time dependent the expression in
the exponential in Eq.\ \eqref{eq:CorrFunInteg} has to be replaced by
an integral, $a_0\imath\varepsilon (t-\tau) \to \int^t_\tau dt' a_0\imath (\varepsilon + \Delta_{\alpha}(t'))$. Consequently, the auxiliary modes take the form,
\begin{equation*}
\chi^{+}_{\alpha p} (t) = [\mu_{\alpha} + \Delta_\alpha (t)] + x_p/\beta.
\end{equation*}
In the following, we will suppress the time-label to facilitate readability of the
equations.

The expansion \eqref{eq:CorrFunAsSum} allows us to find the
equations of motion for the reduced density matrix and the second order \ADO. Moreover, we will
show that one can also obtain an equation of motion for the fourth-order \ADO, which eventually
gives corrections to the second order equations, but also describes new processes.

\subsection{Equations of motion for the second order ADOs} 
Using the expansion of the correlation functions given by Eqs.\ \eqref{eq:CorrFunAsSum} 
we may rewrite the correlation super-operators \eqref{eq:corrsupop} as follows,
\begin{subequations}\label{eq:corrsupopexp}\begin{align}
    \SO{C}^{(2)}_{\mathbf{a} \mathbf{b}}(t,\tau) \bdot = {} &
    \frac{1}{2} \Gamma_{\mathbf{a} \mathbf{b}} \COMM{
      S_{\mathbf{b}}(\tau) }{\bdot} \delta(t-\tau)  
    \nonumber\\ &
    - \sum_p \SO{C}^{(2)}_{\mathbf{a} \mathbf{b} p}(t,\tau) \bdot\;, \\
    \SO{A}^{(2)}_{\mathbf{a} \mathbf{b}}(t,\tau) \bdot = {} &
    \frac{1}{2} \Gamma_{\mathbf{a} \mathbf{b}} \ACOMM{
      S_{\mathbf{b}}(\tau) }{\bdot} \delta(t-\tau) 
    \nonumber\\ &
    - \sum_p \SO{A}^{(2)}_{\mathbf{a} \mathbf{b} p}(t,\tau) \bdot\;,    
\end{align}\end{subequations}
where the sum runs over the auxiliary modes.
The partial super-operators have a very simple structure,
\begin{subequations}\label{eq:corrsupopexpterm}
\begin{align}
    \SO{C}^{(2)}_{\mathbf{a} \mathbf{b} p}(t,\tau) \bdot & = 
    C^{(2)}_{\mathbf{a} \mathbf{b} p}(t,\tau) \ACOMM{ S_{\mathbf{b}}(\tau) }{\bdot}\;, \\
    \SO{A}^{(2)}_{\mathbf{a} \mathbf{b} p}(t,\tau) \bdot & =
    C^{(2)}_{\mathbf{a} \mathbf{b} p}(t,\tau)
    \COMM{ S_{\mathbf{b}}(\tau) }{\bdot} \;,
\end{align}
\end{subequations}
which follows from Eq.~\eqref{eq:CorrFunTerm}. 

Using the expanded super-operators \eqref{eq:corrsupopexp} in
Eq.~\eqref{eq:Def2ndAuxOp} readily leads to an expansion of
the second order \ADO{}s in terms of 
partial operators,
\begin{equation}
  \Pi^{(2)}_{\mathbf{a}}(t) =
   \frac{1}{4}\sum\limits_{\mathbf{b}} 
  \Gamma_{\mathbf{a} \mathbf{b}} \COMM{ S_{\mathbf{b}} }{\RDM(t)}
  - \sum_p \Pi^{(2)}_{\mathbf{a} p}(t) \;.
  \label{eq:ExpPi2}
\end{equation}
The partial operators $\Pi^{(2)}_{\mathbf{a} p}(t)$ are similar to
the definition (\ref{eq:Def2ndAuxOp}) but with
$\SO{C}^{(2)}_{\mathbf{a}\mathbf{b}}$ replaced by
$\SO{C}^{(2)}_{\mathbf{a}\mathbf{b} p}$.
Their equations of motion define the time evolution of 
$\Pi^{(2)}_{\mathbf{a}}(t)$.
These equations read
\begin{align}
  \label{eq:ExpPi2EOM}
  \imath \frac{\partial}{\partial t} \Pi^{(2)}_{\mathbf{a} p}(t) =&
  - \; \sum\limits_{\mathbf{b}} \frac{1}{\beta}\Gamma_{\mathbf{a} \mathbf{b}} \ACOMM{ S_{\mathbf{b}} }{\RDM(t)} \notag \\
  &+ \COMM{ H_{\rm S}(t) }{\Pi^{(2)}_{\mathbf{a} p}(t)} - \chi_{\mathbf{a} p} \Pi^{(2)}_{\mathbf{a} p}(t)
\,,
\end{align}
whereby each of them contains one auxiliary mode $\chi_{\mathbf{a} p}$. 
The initial condition can be found from the definition of the 
partial operators, $\Pi^{(2)}_{\mathbf{a} p}(t_0) = 0$.

\subsection{Equations of motion for the fourth-order ADOs} 
Since the fourth-order \ADO{}s [Eq.\ \eqref{eq:Def4thAuxOp}] are given
in terms of the super-operator $\SO{C}^{(4)}$, we first need an expansion
for these objects. To this end we apply Eqs.\ \eqref{eq:corrsupopexp} in two steps.
First we expand only correlation super-operators containing time
label $t$. This gives
\def\wideornot#1#2{#2}
\wideornot{\begin{widetext}\begin{align}
  \SO{C}^{(4)}_{\mathbf{a} \mathbf{c} \mathbf{d} \mathbf{b}}(t,\tau_1,\tau_2,\tau) \bdot
  = {} & \SO{A}^{(2)}_{\mathbf{c} \mathbf{d}}(\tau_1,\tau_2) \frac{1}{2} \Gamma_{\mathbf{a} \mathbf{b}} \COMM{ S_{\mathbf{b}}(\tau) }{\bdot} \delta(t-\tau)
  - \frac{1}{2} \Gamma_{\mathbf{a} \mathbf{d}} \ACOMM{
    S_{\mathbf{d}}(\tau_2) }{\SO{C}^{(2)}_{\mathbf{c}
      \mathbf{b}}(\tau_1,\tau) \bdot} \delta(t-\tau_2) \nonumber\\
  &\quad- \sum_p \SO{C}^{(4)}_{\mathbf{a} \mathbf{c} \mathbf{d} \mathbf{b} p}(t,\tau_1,\tau_2,\tau) \bdot
\label{eq:4thcorrsupopexp}\end{align}\end{widetext}}{\begin{align}
 &\SO{C}^{(4)}_{\mathbf{a} \mathbf{c} \mathbf{d} \mathbf{b}}(t,\tau_1,\tau_2,\tau) \bdot\notag\\
 &= \SO{A}^{(2)}_{\mathbf{c} \mathbf{d}}(\tau_1,\tau_2) \frac{1}{2} \Gamma_{\mathbf{a} \mathbf{b}} \COMM{ S_{\mathbf{b}}(\tau) }{\bdot} \delta(t-\tau)\notag\\
 &\quad- \frac{1}{2} \Gamma_{\mathbf{a} \mathbf{d}} \ACOMM{
   S_{\mathbf{d}}(\tau_2) }{\SO{C}^{(2)}_{\mathbf{c}
     \mathbf{b}}(\tau_1,\tau) \bdot} \delta(t-\tau_2) \notag\\
 &\quad- \sum_p \SO{C}^{(4)}_{\mathbf{a} \mathbf{c} \mathbf{d} \mathbf{b} p}(t,\tau_1,\tau_2,\tau) \bdot
\label{eq:4thcorrsupopexp}\end{align}}
with
\begin{align}
	\SO{C}^{(4)}_{\mathbf{a} \mathbf{c} \mathbf{d} \mathbf{b} p}(t,\tau_1,\tau_2,\tau) = {} &
		 \SO{A}^{(2)}_{\mathbf{c} \mathbf{d}}(\tau_1,\tau_2) \SO{C}^{(2)}_{\mathbf{a} \mathbf{b} p}(t,\tau)  \notag \\
		&- \SO{A}^{(2)}_{\mathbf{a} \mathbf{d} p}(t,\tau_2) \SO{C}^{(2)}_{\mathbf{c} \mathbf{b}}(\tau_1,\tau) \,.
\end{align}
With these results one can expand the fourth-order {\ADO} given by Eq.~\eqref{eq:Def4thAuxOp}. Notice that
the nested time integrals in \eqref{eq:Def4thAuxOp} are such, that the terms in Eq.~\eqref{eq:4thcorrsupopexp} containing 
a $\delta$-function vanish. Therefore, one has 
\begin{subequations}\begin{align}
\Pi^{(4)}_{\mathbf{a}}(t) = {} & \sum_p \Pi^{(4)}_{\mathbf{a} p}(t)
\label{eq:ExpPi4}\\
\Pi^{(4)}_{\mathbf{a} p}(t) = {} & -\sum\limits_{\mathbf{b}, \mathbf{c}, \mathbf{d}} 		
\int\limits_{t_0}^t d\tau \int\limits_{\tau}^t d\tau_1 \int\limits_{\tau}^{\tau_1} d\tau_2 \;
\SO{U}_{\rm S}(t)\times \notag\\
&\ACOMM{S_{\mathbf{c}}(\tau_1)}{
  \SO{C}^{(4)}_{\mathbf{a} \mathbf{c} \mathbf{d} \mathbf{b} p}(t,\tau_1,\tau_2,\tau) \RDM(\tau) 
} \;.
\label{eq:4thAuxOpExpDef}
\end{align}\end{subequations}
In analogy to the second order case one can give an equation of motion for $\Pi^{(4)}_{\mathbf{a} p}$,
\begin{align}
	\imath \frac{\partial}{\partial t} \Pi^{(4)}_{\mathbf{a} p}(t) =&
		-\imath \sum\limits_{\mathbf{c}} \ACOMM{S_{\mathbf{c}}}{ \Phi^{(4)}_{\mathbf{a} \mathbf{c} p}(t) } \notag \\
		&+ \COMM{ H_{\rm S}(t) }{\Pi^{(4)}_{\mathbf{a} p}(t)} - \chi_{\mathbf{a} p} \Pi^{(4)}_{\mathbf{a} p}(t) \;,
\label{eq:ExpPi4EOM}
\end{align}
which contains only one auxiliary mode.
The last two terms result from the time-derivative of the system propagator and the reservoir correlation function, respectively.
The first term, denoted by $\Phi^{(4)}_{\mathbf{a} \mathbf{c} p}(t)$, comes from the derivative of the integrals and hence is given by
\begin{align}
  \Phi^{(4)}_{\mathbf{a} \mathbf{c} p}(t) &= \sum\limits_{\mathbf{b}, \mathbf{d}} 		
        \int\limits_{t_0}^t d\tau \int\limits_{\tau}^{t} d\tau_2 \times\notag\\			
				&\qquad\quad \SO{U}_{\rm S}(t) \SO{C}^{(4)}_{\mathbf{a} \mathbf{c} \mathbf{d} \mathbf{b} p}(t,t,\tau_2,\tau) \RDM(\tau) 
			\;.
\end{align}
The equation of motion is supplemented by the initial condition found from Eq.~\eqref{eq:4thAuxOpExpDef} to be
$\Pi^{(4)}_{\mathbf{a} p}(t_0) = 0$.
Now we expand the remaining two correlation super-operators in $\SO{C}^{(4)}$ formerly involving time label $\tau_1$,
\wideornot{\begin{widetext}\begin{align}
      \SO{C}^{(4)}_{\mathbf{a} \mathbf{c} \mathbf{d} \mathbf{b} p}(t,t,\tau_2,\tau) \bdot
      = {} & \frac{1}{2} \Gamma_{\mathbf{c} \mathbf{d}} \ACOMM{ S_{\mathbf{d}}(t) }{\SO{C}^{(2)}_{\mathbf{a} \mathbf{b} p}(t,\tau) \bdot} \delta(t-\tau_2) 
      - \SO{A}^{(2)}_{\mathbf{a} \mathbf{d} p}(t,\tau_2) \frac{1}{2} \Gamma_{\mathbf{c} \mathbf{b}} \COMM{ S_{\mathbf{b}}(\tau) }{\bdot} \delta(t-\tau) \notag \\
      &- \sum_q \SO{C}^{(4)}_{\mathbf{a} \mathbf{c} \mathbf{d} \mathbf{b} p q}(t,t,\tau_2,\tau) \bdot \;.
      \label{eq:4psupercorrexp}\end{align}\end{widetext}}{\begin{align}
    &\SO{C}^{(4)}_{\mathbf{a} \mathbf{c} \mathbf{d} \mathbf{b} p}(t,t,\tau_2,\tau) \bdot\notag\\
    &= \frac{1}{2} \Gamma_{\mathbf{c} \mathbf{d}} \ACOMM{ S_{\mathbf{d}}(t) }{\SO{C}^{(2)}_{\mathbf{a} \mathbf{b} p}(t,\tau) \bdot} \delta(t-\tau_2) \notag\\
    &\quad - \SO{A}^{(2)}_{\mathbf{a} \mathbf{d} p}(t,\tau_2) \frac{1}{2} \Gamma_{\mathbf{c} \mathbf{b}} \COMM{ S_{\mathbf{b}}(\tau) }{\bdot} \delta(t-\tau) \notag \\
    &\quad- \sum_q \SO{C}^{(4)}_{\mathbf{a} \mathbf{c} \mathbf{d} \mathbf{b} p q}(t,t,\tau_2,\tau) \bdot \;.
    \label{eq:4psupercorrexp}\end{align}}
Inserting this expression into $\Phi^{(4)}_{\mathbf{a} \mathbf{c} p}(t)$, one observes that due to the integral boundaries the term containing the delta function $\delta(t-\tau)$
vanishes and one is left with 
\begin{equation}
  \Phi^{(4)}_{\mathbf{a} \mathbf{c} p}(t) =
  \frac{1}{4}\sum\limits_{\mathbf{d}} 		
  \Gamma_{\mathbf{c} \mathbf{d}} \ACOMM{ S_{\mathbf{d}} }{ \Pi^{(2)}_{\mathbf{a} p}(t) }					
  - \sum_q \Phi^{(4)}_{\mathbf{a} \mathbf{c} p q}(t)\;,
\end{equation}
and
\begin{equation}\label{eq:twopartauxopdef}
  \Phi^{(4)}_{\mathbf{a} \mathbf{c} p q}(t) =
  \sum\limits_{\mathbf{b}, \mathbf{d}} 		
  \int\limits_{t_0}^t d\tau \int\limits_{\tau}^{t} d\tau_2 \;			
  \SO{U}_{\rm S}(t) \SO{C}^{(4)}_{\mathbf{a} \mathbf{c} \mathbf{d} \mathbf{b} p q}(t,t,\tau_2,\tau) \RDM(\tau) 
  \;.
\end{equation}
Obviously one gets two qualitatively different contributions to the equation of motion for $\Pi^{(4)}$. The first one being proportional
to $\Pi^{(2)}$ contains only one excitation (additional electron or hole) in the reservoirs and gives corrections to the equation
of motion for $\Pi^{(2)}$. The second contribution, given by $\Phi^{(4)}_{\mathbf{a} \mathbf{c} p q}$, describes two excitations in the reservoirs, i.\,e.~the onset of
genuine {\it co-tunneling}. 

In summary, we write for the equation of motion of the fourth-order partial operators,
\begin{widetext}
\begin{equation}
  \label{eq:4thAuxOpEOM}
  \imath \frac{\partial}{\partial t} \Pi^{(4)}_{\mathbf{a} p}(t) = {} 
  -\imath \sum\limits_{\mathbf{c}} 
  \ACOMM{S_{\mathbf{c}}}{
    \frac{1}{4}\sum\limits_{\mathbf{d}} \Gamma_{\mathbf{c} \mathbf{d}} \ACOMM{ S_{\mathbf{d}} }{ \Pi^{(2)}_{\mathbf{a} p}(t) }	
 - \sum_q \Phi^{(4)}_{\mathbf{a} \mathbf{c} p q}(t)
  } 
  + \COMM{ H_{\rm S}(t) }{\Pi^{(4)}_{\mathbf{a} p}(t)} - \chi_{\mathbf{a} p} \Pi^{(4)}_{\mathbf{a} p}(t) \;.
\end{equation}
In order to get a complete description we also need the time
evolution of $\Phi^{(4)}_{\mathbf{a} \mathbf{c} p q}$, which can be obtained
by using again a differential equation,
\begin{equation}
    \imath \frac{\partial}{\partial t} \Phi^{(4)}_{\mathbf{a} \mathbf{c} p q}(t) =
    \imath \sum\limits_{\mathbf{d}} \left(
      \SO{A}^{(2)}_{\mathbf{c} \mathbf{d} q}(t,t+0) \Pi^{(2)}_{\mathbf{a} p}(t)
      -\SO{A}^{(2)}_{\mathbf{a} \mathbf{d} p}(t,t+0) \Pi^{(2)}_{\mathbf{c} q}(t)
    \right)
    + \COMM{ H_{\rm S}(t) }{\Phi^{(4)}_{\mathbf{a} \mathbf{c} p q}(t)} - \left( \chi_{\mathbf{a} p} + \chi_{\mathbf{c} q} \right) \Phi^{(4)}_{\mathbf{a} \mathbf{c} p q}(t) \;.
\end{equation}\end{widetext}
The initial condition is easily obtained, $\Phi^{(4)}_{\mathbf{a} \mathbf{c} p q}(t_0) = 0$. Using Eqs.~\eqref{eq:corrsupopexpterm}
and \eqref{eq:twopartauxopdef} one can also show that $\Phi^{(4)}_{\mathbf{a} \mathbf{c} p q}$ is a traceless operator, 
\begin{equation}\label{eq:phi4trless}
  \Tr_{\rm S} \left\{ \Phi_{\mathbf{a} \mathbf{c} p q}(t) \right\} = 0 \;.
\end{equation}
This property will be important in the next section, when we consider expectation values
for example to calculate the electric current.

\subsection{Electric Current}\label{sec:CurrentADO}
So far we have given a closed set of differential equations for
the reduced density operator and newly introduced partial
operators.
By means of these quantities one can calculate the time evolution of any observable of the system. Moreover,
one can also compute quantities like the time-dependent electric current, which do not belong to
the system state space alone.

Up to fourth order the electric current \eqref{eq:CurrPartCum} becomes,
\begin{align}
  \label{eq:Curr4thOrder}
  J^{(4)}_\alpha (t) = {} &  {2 e\;} \Re \left\{
    \sum\limits_{m}
    \Tr_{\rm S} \left[ c_m \Pi^{(2, +)}_{\alpha m} (t) \right]
  \right. 
  \nonumber\\ &\qquad
  -\left.\sum\limits_{m}
    \Tr_{\rm S} \left[ c_m \Pi^{(4, +)}_{\alpha m} (t) \right]
\right\}\;,
\end{align}
where the definition of the \ADO{}s given by Eqs.~\eqref{eq:Def2ndAuxOp} and 
(\ref{eq:Def4thAuxOp}) has been used.
Using the expanded forms of the second and fourth order
operators, cf.\ Eqs.~(\ref{eq:ExpPi2}) and (\ref{eq:ExpPi4}), one finds
\begin{align}
  &\Tr_{\rm S} \left[ c_m \Pi^{(2, +)}_{\alpha l} (t) - c_m \Pi^{(4, +)}_{\alpha l} (t) \right] \notag\\
  &= \sum\limits_{j} \frac{1}{4} \left( \bar{\RDM}_{mj} (t) - \RDM_{mj} (t)\right) \Gamma_{\alpha,jl}  \notag \\
  &\quad-\sum\limits_p \Tr_{\rm S} \left[ c_m \Pi^{(2, +)}_{\alpha lp} (t) 
    + c_m \Pi^{(4, +)}_{\alpha lp} (t) \right] \;,
\end{align}
where $\RDM_{mj}(t) = \Tr_{\rm S} \big\{ c^\dagger_j c_m \RDM(t) \big\}$ is the one-particle and
$\bar{\RDM}_{mj}(t) = \Tr_{\rm S} \big\{ c_m c^\dagger_j \RDM(t) \big\}$ is the one-hole density matrix of the system.
The current is therefore given in terms of the reduced one-particle density matrix and the auxiliary operators.
For the last terms involving the sum over expanded \ADO{}s one might now use the equation of motions derived in
the previous section, cf.\ Eqs.\ (\ref{eq:ExpPi2EOM}) and
(\ref{eq:ExpPi4EOM}). 

In an occupation number representation the traces in the
current expression as well as the equations of motion can be
solved numerically for an arbitrary device Hamiltonian
(\ref{eq:DevHamilOp}) including interactions.
As an example we will show results for an interacting double
quantum dot system in Sect.~\ref{sec:Appl}.

\section{Infinite orders}\label{sec:InfiniteDensityMatrix}
\noindent
In the previous section we have considered the cumulant
expansion up to fourth order. Naturally, the question arises how
the higher orders influence the dynamics. In principle,
one has to sum up all orders to get an exact description.
This problem has been tackled, for example, diagrammatically
\cite{scsc94} or by using a path-integral formalism
\cite{jizh+08}. Moreover, the connection of cumlant expansions
obtained from the projection operator technique to the diagrammatic
expansion has been discussed recently \cite{ti08}. Other approaches 
consequently use many-body states to derive equations of motion
\cite{pewa05,pela+07,esga09}.

In the following, we will first show for non-interacting electrons  that summing up all orders with one
excitation yields a QME, which resembles the equations of motion
from the well-known NEGF formalism, cf.\ 
appendix \ref{sec:NEGF}.
It contains level broadening of the system states due to
coupling to the reservoirs.
Further, we adopt the path-integral formalism \cite{jizh+08} and device a propagation 
scheme for systems with electron-electron interactions similar to the one developed in the
previous section for the WBL. We restrict the discussion
to single-electron excitations in the reservoirs only.

\subsection{Noninteracting Electrons} \label{sec:effauxopeom}
Neglecting electron-electron interactions we can compare
the expressions obtained from the QME and the NEGF formalism. For example, the 
electric current up to fourth order is given by
\begin{align}
  \label{eq:QMECurr4thOrder}
  J^{(4)}_\alpha (t) = {} &  {2 e\;} \Re \sum\limits_{m} \left\{
    \frac{1}{4}\sum_l \left( \bar{\RDM}_{m l} (t) - \RDM_{m l}(t)\right) 
    \Gamma_{\alpha, l m}  \right.\notag \\	
  &\left. \quad-\sum\limits_p \left( P^{(2, +)}_{\alpha p;m m} (t) + P^{(4, +)}_{\alpha p;m m} (t) \right)
  \right\}\;,
\end{align}
and therefore in terms of single-particle quantities $\RDM_{ml}(t)$ and 
$P^{(2n,\pm)}_{\alpha;ml}(t) =  
\Tr_{\rm S} \left[ c_m \Pi^{(2n,\pm)}_{\alpha l} (t) \right]$ with $n=1,2$.
In order to get a complete description one has to calculate the equations 
of motion for these matrices, which
may be obtained directly from the respective QME.

Considering the matrices $P^{(2n,\pm)}_{\alpha;ml}$ we find from Eq.\ \eqref{eq:ExpPi2EOM}
that their time derivative for the second-order case is given by
\begin{align}
  \imath \frac{\partial}{\partial t} P^{(2, +)}_{\alpha p;ml} (t)
  =& -\frac{1}{\beta}\sum\limits_{j} \left[ \bar{\RDM}_{mj} (t) 
    + \RDM_{mj} (t)\right] \Gamma_{\alpha j l} \notag\\
   &+ \sum\limits_{j} h_{mj} P^{(2, +)}_{\alpha p;jl} (t) 
   - \chi^+_{\alpha p} P^{(2, +)}_{\alpha p;ml} (t) \;
\label{eq:2ndCurrEOM}
\end{align}
with $h_{mj}=\varepsilon_m\delta_{mj}+V_{mj}$ as an
abbreviation for the device Hamiltonian (\ref{eq:DevHamilOp}).
The fourth-order contribution is obtained from Eq.\ \eqref{eq:4thAuxOpEOM} and reads
\begin{align}\label{eq:4thCurrEOM}
  \imath \frac{\partial}{\partial t} P^{(4, +)}_{\alpha p;ml} (t)
  =& -\imath \frac{1}{2} \sum\limits_{\alpha' j}  \Gamma_{\alpha' mj} P^{(2, +)}_{\alpha p;jl} (t) \notag\\
  &  +\sum\limits_{j} h_{mj} P^{(4, +)}_{\alpha p;jl} (t) - \chi^+_{\alpha p} P^{(4, +)}_{\alpha p;ml} (t) \notag\\
  &  + \imath \sum\limits_{q} \Tr_{\rm S} \left[\Phi^{(4,+-)}_{\alpha l\;\alpha' m,p q}(t) \right] \;.
\end{align}
whereby the term $\Tr_{\rm S} \left\{ \Phi^{(4,+-)}_{\alpha m\;\alpha' l,pq}(t) \right\}$ 
is easily seen to vanish by using Eqs.~\eqref{eq:twopartauxopdef} and 
\eqref{eq:corrsupopexp}.

Comparing Eq.~\eqref{eq:EffCurr4thOrder} with the exact NEGF result given by Eqs.~\eqref{eq:GenCurrent3} and
\eqref{eq:AuxCurrMatDef} suggests that our results constitute the first terms of an 
expansion eventually leading to 
\begin{equation}
  P^{(+)}_{\alpha p;ml}(t) 
  = \sum\limits^\infty_{n=1} P^{(2 n, +)}_{\alpha p;ml}
  = -\left[\boldgreek{\Pi}_{\alpha p}(t,t)\right]_{ml} \;,
\end{equation}
where each term $P^{(2 (n+1), +)}$ gives rise to corrections for the lower-order term $P^{(2 n, +)}$. Indeed, one 
sees that generalizing Eq.~\eqref{eq:4thCurrEOM} to
\begin{align}
  \imath \frac{\partial}{\partial t} P^{(2n, +)}_{\alpha p;m l} (t)
  = {} & - \frac{\imath}{2} \sum\limits_{\alpha' j}  
  \Gamma_{\alpha', m j} P^{(2n-2, +)}_{\alpha p;j l} (t) 
  \notag\\ &  +\sum\limits_{j} h_{m j} P^{(2n, +)}_{\alpha p;j l} (t) 
  \notag\\ &- \chi^+_{\alpha p} P^{(2n, +)}_{\alpha p;m i} (t) \label{eq:2nthCurrEOM}
\end{align}
with $n \ge 2$ yields the correct equation of motion for $P^{(+)}_{\alpha p;ml}$. 
By comparing with Eq.\ \eqref{eq:QMECurr4thOrder} the current is given by
\begin{align}
  \label{eq:EffCurr4thOrder}
  J_\alpha (t) = {} &  {2 e\;} \Re \sum\limits_{m} \left\{
    \frac{1}{4}\sum_l \left( \bar{\RDM}_{m l} (t) - \RDM_{m l}(t)\right) 
    \Gamma_{\alpha, l m}  \right.\notag \\	
  &\qquad\left. \quad-\sum\limits_p  P^{(+)}_{\alpha p;m m} (t) 
  \right\}\;.
\end{align}
This expression is identical to the respective result obtained from the NEGF
formalism, cf.\ appendix \ref{sec:NEGF}.

\subsection{Interacting electrons}\label{sec:CohOpHier}
Inspecting Eq.~\eqref{eq:4thAuxOpEOM}, which led to Eq.~\eqref{eq:4thCurrEOM} in the
first place, one is inclined to conclude that for arbitrary order $n\ge 2$ one has
to use a modified EOM,
\begin{align}
  \label{eq:2nthAuxOpEOM}
  \imath \frac{\partial}{\partial t} \Pi^{(2n)}_{\mathbf{a} p}(t) 
  = {} &- \frac{\imath}{4} \sum\limits_{\mathbf{c} \mathbf{d}} \Gamma_{\mathbf{c} \mathbf{d}}
      \ACOMM{ S_{\mathbf{c}} }{
          \ACOMM{ S_{\mathbf{d}} }{ \Pi^{(2n-2)}_{\mathbf{a} p}(t) }	
      } \notag \\
  &+ \COMM{ H_{\rm S}(t) }{\Pi^{(2n)}_{\mathbf{a} p}(t)} - \chi_{\mathbf{a} p} \Pi^{(2n)}_{\mathbf{a} p}(t) \nonumber  \\
  &+\imath \sum_q \ACOMM{ S_{\mathbf{c}} }{
    \Phi^{(2 n)}_{\mathbf{a} \mathbf{c} p q}(t)
  } \;. 
\end{align}
This equation generalizes Eq.~\eqref{eq:4thAuxOpEOM} and leads to the correct form
given by Eq.\ \eqref{eq:2nthCurrEOM}. Therefore, this generalized EOM yields the
exact dynamics for vanishing electron-electron interactions. 

For a rigorous derivation one has to sum all higher-order cumulants.
Instead of explicitly summing all higher-order cumulants in Eqs.~\eqref{eq:TNLQMEIP} and 
\eqref{eq:CurrPartCum} we use the hierarchy of equations of motion for the \ADO{}s
derived elsewhere \cite{jizh+08}. Since we are interested in the WBL we have to modify the equations
given therein. In particular the term containing the $\delta$-function in the expansion
for the correlation functions given in Eqs.~(\ref{eq:CorrFunAsSum}) leads to a substantial
simplification. The equation of motion for the reduced density matrix is almost identical
to Eq.~\eqref{eq:EOMTNL4thQME},
\begin{equation}
  \imath \frac{\partial}{\partial t} \RDM(t) 
  = \COMM{H_{\rm S}(t)}{ \RDM(t) }
   -\imath \sum\limits_{\mathbf{a}} \COMM{S_{\mathbf{a}}}{
     \AuxPsi_{\mathbf{a}}(t)
       }\;.
   \label{eq:HEOMRDM}
\end{equation}
The new operators $\AuxPsi_{\mathbf{a}}$ involve all orders $\Pi^{(2n)}$. Using
the expansion of the correlation functions given in Eqs.~(\ref{eq:CorrFunAsSum}) we obtain
the following expansion
\begin{align}
\AuxPsi_{\mathbf{a}}(t) 
  =& \sum\limits^\infty_{n=1} \Pi^{(2n)}_{\mathbf{a}} (t) \notag \\
  =& \sum\limits_{\mathbf{b}} \frac{1}{4} \Gamma_{\mathbf{a} \mathbf{b}} \COMM{S_{\mathbf{b}}}{\RDM(t)}
       -\sum\limits_p \AuxPsi_{\mathbf{a} p} (t)\;.
  \label{eq:EOMAuxDef}
\end{align}
which reminds of Eq.~\eqref{eq:ExpPi2}. The electric current is then calculated from the 
single-electron \ADO{}s only \cite{jizh+08}
\begin{equation}\label{eq:HEOMCurrent}
  J_\alpha (t) = 2 e\; \Re \sum\limits_m \Tr_{\rm S} \left[ c_m \AuxPsi^+_{\alpha m} (t) \right] \;,
\end{equation}
which resembles Eq.~\eqref{eq:CurrPartCum}.

The first level of the hierarchy is given by the EOM for 
the auxiliary operator $\AuxPsi_{\mathbf{a} p}$, which
describes the single-electron partial \ADO{} with
one reservoir excitation denoted by auxiliary mode $p$. Its equation of motion reads
\begin{align}
  \imath \frac{\partial}{\partial t} \AuxPsi_{\mathbf{a} p} (t)
  = {} &-\sum\limits_{\mathbf{b}} \frac{1}{\beta} \Gamma_{\mathbf{a} \mathbf{b}}  \ACOMM{S_{\mathbf{b}}}{ \RDM(t) } \notag \\
    &+ \COMM{ H_{\rm S}(t) }{ \AuxPsi_{\mathbf{a} p} (t) }
    - \chi_{\mathbf{a} p} \AuxPsi_{\mathbf{a} p} (t) \notag\\
    &-\imath \sum\limits_{\mathbf{c} \mathbf{d}} \frac{1}{4} \Gamma_{\mathbf{c} \mathbf{d}}
          \ACOMM{S_{\mathbf{c}}}{ \ACOMM{S_{\mathbf{d}}}{ \AuxPsi_{\mathbf{a} p} (t)}} \notag \\
    &+ \imath \sum\limits_{\mathbf{c} q} \ACOMM{ S_{\mathbf{c}}}{ \AuxPsi_{\mathbf{a} \mathbf{c} p q} (t) } \;,
    \label{eq:HEOMAux}
\end{align}
where we recognize the structure of Eqs.~\eqref{eq:ExpPi2EOM} and \eqref{eq:4thAuxOpEOM}.
Obviously, Eq.\ \eqref{eq:HEOMAux} contains couplings to the 0th and 2nd level
of the hierarchy, which are given by the first and the last term, respectively. 
The operator $\AuxPsi_{\mathbf{a} \mathbf{c} p q} = \sum^\infty_{n=2} \Phi^{(2n)}_{\mathbf{a} \mathbf{c} p q}$ surmounts all contributions with two auxiliary modes. The
second level of the hierarchy is then given by the EOM for $\AuxPsi_{\mathbf{a} \mathbf{c} p q}$. In order to truncate the hierarchy, we set $\AuxPsi_{\mathbf{a} \mathbf{c} p q}=0$,
which does not influence the results for non-interacting electrons (see Eq.\ \eqref{eq:4thCurrEOM}). This Ansatz is not unique, however, it is very convenient since it preserves
the linearity of the EOMs. The differential equations \eqref{eq:HEOMRDM}, \eqref{eq:HEOMAux} and the expansion \eqref{eq:EOMAuxDef} provide a closed description of
the evolution of the reduced density operator. These equations will be denoted as
{\it effective quantum master equations} in the following. As it was shown above, these
equations are sufficient to calculate single-particle quantities like the electron
current for vanishing electron-electron interactions. This is true for non-interacting electrons, but also for a very large interaction strength, where only one
electron can enter in the system.
In contrast to the finite order equations (second and fourth order QMEs), the effective
QME consistently treats the broadening of the electronic levels due to back-action of the electron reservoirs on the system, which is also included
in the NEGF formalism. The correct handling of the level-broadening is a necessary 
ingredient for the description of electron transport as we will show in the next
section for a specific example.

\section{Numerical Results: Double Quantum Dot}\label{sec:Appl}
\noindent
In this section we will implement the propagation schemes developed in the previous sections. In particular, we will investigate the influence of higher-order terms
in the QME and compare the results with other methods.
To this end we consider
a double quantum dot (DQD) system, where the two dots are coupled in series and each dot 
is connected to an electron reservoir. Over the past years this system has become a paradigm for electron transport through a nontrivial quantum system \cite{pela+07,scki+09,crsa10a}.
Moreover, the advent
of explicitly time-resolved experiments, such as coherent control investigations \cite{hafu+03,fuha+06}, allows to test the suggested theories.

\subsection{Setup}
In following we will consider only spinless electrons, which simplifies the calculations 
to some extend, but is not mandatory otherwise. The Hamiltonian of the reservoirs and the
coupling of the DQD to the reservoirs is given in Eqs.~\eqref{eq:ResHamilOp} and 
\eqref{eq:TunnHam}. The Hamiltonian describing the DQD is explicitly given by
\def\eps{\varepsilon}\def\l#1{#1_{\ell}}\def\r#1{#1_\mathrm{r}}
\begin{align}
    H_{\rm S} = {} & \eps \, \l{c^\dagger} \l{c} 
  - \eps \, \r{c^\dagger} \r{c} 
  \notag\\ &+ V (\l{c^\dagger}\r{c} + \r{c^\dagger}\l{c})
  + U \l{c^\dagger}\l{c}\r{c^\dagger}\r{c} \;,
\label{eq:QMEAppDQDHam}\end{align}
where the localized states of the left and right dot are denoted
by ``$\ell$'' and ``r'', respectively. The first three terms in Eq.\ \eqref{eq:QMEAppDQDHam}
describe an effective two-level system, which is characterized
by the interdot tunnel coupling $V$ and energies $+\eps$ and $-\eps$, respectively.
The interdot Coulomb repulsion-strength is denoted by $U$.
The eigenenergies of the Hamiltonian \eqref{eq:QMEAppDQDHam} are
\begin{equation}\label{eq:APPEigEn}
E_0 = 0,\quad
E_1^\pm =\pm \sqrt{\eps^2 {+} V^2},\quad
E_2 = U.
\end{equation}
 These energies
correspond to configurations with zero ($E_0$), one ($E_1^\pm$) and two ($E_2$) electrons in the DQD, respectively. 

Since the two dots are coupled in series the level-width functions in the WBL contain one non-zero element,
\begin{gather*}
  \boldgreek{\Gamma}_{\rm L} = 
  \left( \begin{array}{cc}
      \Gamma/2 & 0 \\
      0 & 0 
    \end{array}
  \right), \quad\boldgreek{\Gamma}_{\rm R} =
  \left( \begin{array}{cc}
      0 & 0 \\
      0 & \Gamma/2
    \end{array}
  \right)\;.
\end{gather*}
Here, we have assumed a symmetric coupling of the left and right dot to their respective
reservoir. Following Ref.~\cite{pewa05} we assume a symmetric voltage drop across the system, $\mu_{\rm L}=-\mu_{\rm R}=V_{\rm bias}/2$ und $\eps=\Gamma/2$.

The QMEs are conveniently represented using the occupation number basis consisting
of four states: $\ket{0}, \ket{\ell}, \ket{\rm r}$ and $\ket{2}$. Consequently, all 
operators become $4 {\times} 4$ matrices. The matrix equations resulting from the equations 
of motion are propagated using a fourth order Runge-Kutta scheme \cite{prfl+92}. 
The number of auxiliary modes is $N_{\rm F} = 120$. In all cases 
we start the simulation by suddenly coupling the initially empty DQD to the reservoirs at $t=0$. The temperature of the electrons in the reservoirs is set to $T = 0.1 \Gamma/k_{\rm B}$.

We consider the second and fourth order QMEs, hereafter referred to as QME2 and QME4, given by Eqs.~\eqref{eq:EOMTNL4thQME}, \eqref{eq:ExpPi2EOM} 
and \eqref{eq:4thAuxOpEOM}, where for the latter we set $\Phi^{(4)}_{\mathbf{a} \mathbf{c},pq} = 0$. 
The QME resulting from the hierarchy will be denoted as effective QME (QMEe). We will also present
results for the NEGF method and the Markovian QME, which are explained in App.\ \ref{sec:NEGF} and \ref{sec:Markov},
respectively.

\subsection{Stationary Current}
Before turning to the transient behavior of the different QMEs we will address the 
stationary current as a function of the bias voltage. To this end we propagate all QMEs as 
described in the previous section up to a final time $t=60 \Gamma^{-1}$. 

\begin{figure}[tbp!]
  \centering
  \includegraphics[scale=0.5]{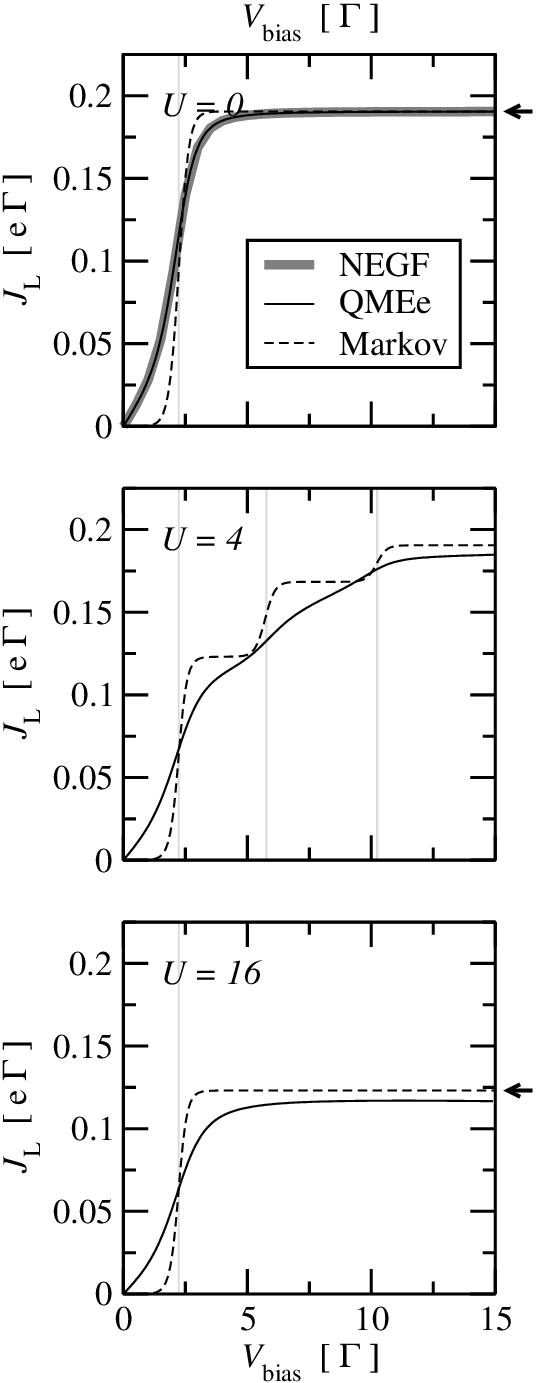}
  \includegraphics[scale=0.5]{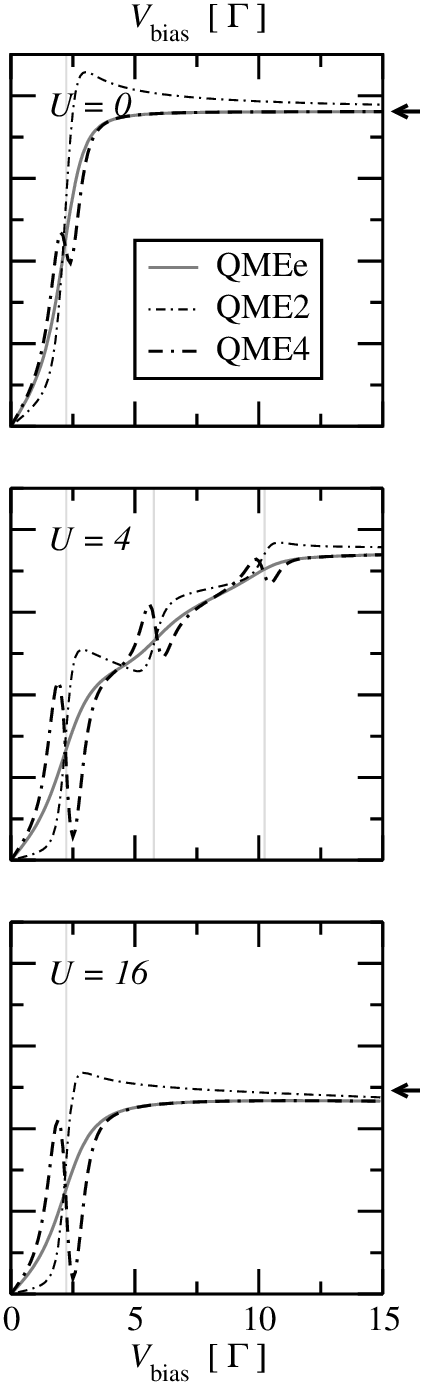}
  \caption{Stationary current $J_{L}$ vs bias voltage $V_{\rm bias}$ for different values of
   the Coulomb repulsion $U$. Left panel: results obtained
	from the effective and the Markovian QME. Right panel: results obtained using the 
	second order and fourth order QME and the effective QME. 
   Vertical lines indicate the transition energies. Arrows show the values obtained 
   independently in the large-bias limit \cite{stna96,gupr96}.
  }\label{fig:AppStatCurr}   
\end{figure}
Figure \ref{fig:AppStatCurr} shows the stationary current for different values of $U$. 
The left column provides a comparison of the effective QME with the Markovian QME. For 
$U=0$ we also show exact results using the NEGF formalism \cite{wija+93} and the
large-bias limit \cite{stna96, gupr96}. The right column gives a comparison of 
QMEe with the second and fourth order counterparts. 

The current displays a single ($U=0,16\Gamma$) or multiple step behavior ($U=4\Gamma$), which is monotonic 
for the effective and Markovian QME, but non-monotonic for the second and fourth order order calculations. 
The position of the steps is best understood in terms of
the eigenenergies \eqref{eq:APPEigEn} of the system Hamiltonian and the allowed 
transitions between them. Transitions are only allowed by
changing the electron number by one, i.e.\ 
either $0\leftrightarrow1^\pm$ or $1^\pm\leftrightarrow2$. 
The steps in Fig.~\ref{fig:AppStatCurr} occur, when a new transition energy, 
$E_{AB}=E_A-E_B$, becomes accessible within the transport window 
$\{\mu_{\rm L},\mu_{\rm R}\} = \{V_{\rm bias}/2, -V_{\rm bias}/2\}$. Therefore,
a step in the current-voltage curve is expected for $V_{\rm bias}/2 = \pm E_{AB}$. The
transition energies are given in Tab.\ \ref{tab:APPTranEn} and the respective
voltages are indicated in Fig.~\ref{fig:AppStatCurr} by vertical lines. Due
to the symmetry of the setup only up to three transitions are visible in Fig.~\ref{fig:AppStatCurr}.
\begin{table}[bp!]
	\begin{ruledtabular}
  	\begin{tabular}[b]{@{\extracolsep{\fill}}lccc}
    & \multicolumn{3}{c}{$E_{AB}$} \\ \cline{2-4}
    & $0\to 1^\pm$ & $1^-\to 2$ & $1^+\to 2$ \\
    \hline
    $U=0$  & $\pm 1.118$  & $1.118$ & $-1.118$ \\
    $U=4$  & $\pm 1.118$  & $5.118$ & $2.882$ \\
    $U=16$ & $\pm 1.118$  & $17.118$ & $14.882$ \\
    \hline
    $U=\infty$  & $\pm 1.118$  & $\infty$ & $\infty$\\
  	\end{tabular}
  	\end{ruledtabular}
  	\caption
    {Transition energies for allowed transitions. The energies are
     calculated by using Eqs.\ \eqref{eq:APPEigEn} with
     $\eps=\Gamma/2$ und $V=\Gamma$.}%
  	\label{tab:APPTranEn}
\end{table}

The Markovian results show clear steps
for all values of $U$. The width of the rising edges is of the order $k_{\rm B} T$. In contrast to that, the effective
QME yields much broader steps. This is a result of the broadening of the energy levels due to the coupling
to the reservoirs, which is not present in the second order description underlying the Markovian approximation.
In the present case the coupling strength to the reservoirs is 
larger than the temperature, $\Gamma/2 > k_{\rm B} T$, and thus the rising of the step is
mainly determined by the tunnel coupling.
For a sufficiently large bias voltage all energy levels are within the transport window regardless 
of being broadened or not. In this limit both methods yield the same results, i.e.\ the
Markovian approximation is sufficient. This is readily seen in Fig.~\ref{fig:AppStatCurr}.
The small deviations for finite $U$ might be attributed to neglecting the
contributions of the two-electron \ADO{}s.

\begin{figure}[tbp!]
  \centering
  \includegraphics[scale=0.7]{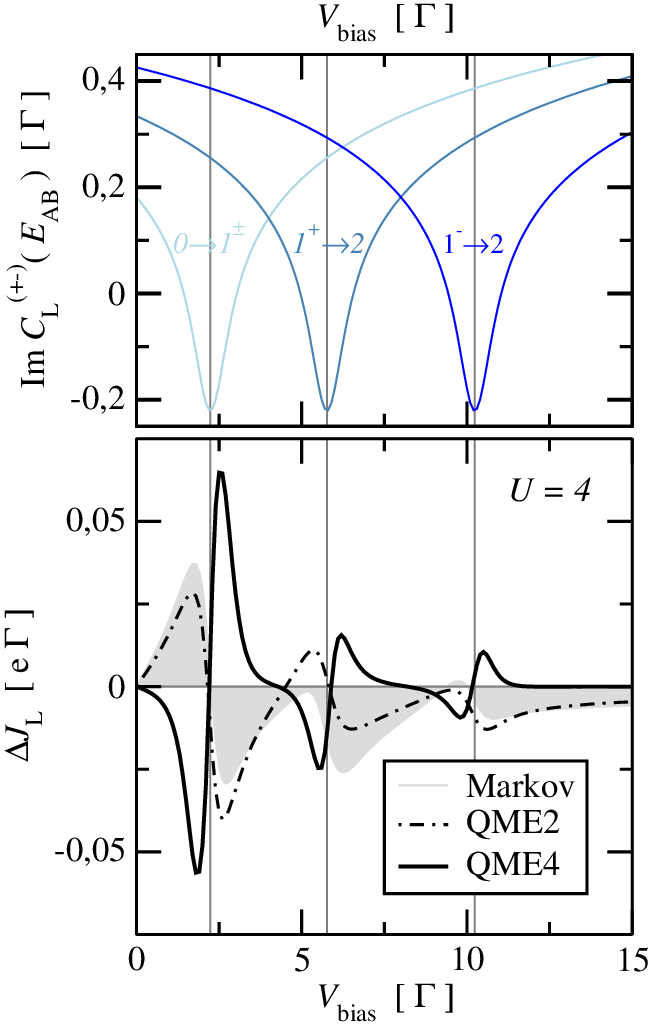}
  \caption{
  	 Upper graph: Laplace transform of the reservoir correlation function $C^{(+-)}_{\rm L}$
	 according to Eq.\ \eqref{eq:MarkovCLTrans}.
	 Lower graph:
	 Stationary current $J_{L}$ vs bias voltage $V_{\rm bias}$ for Coulomb repulsion 
	 $U = 4\Gamma$. Curves show
    the difference with respect to the effective QME results.
    Vertical lines indicate the transition energies. Deviations for
    the Markovian QME are indicated by the gray-shaded areas.}
    \label{fig:AppStatCurrU4}   
\end{figure}

The second and fourth order calculations show prominent features at the transition energies,
which are not expected within the usual picture of sharp energy levels. Figure \ref{fig:AppStatCurrU4} 
shows the deviations of the QME2 and QME4 results from the effective QME calculations for $U=4\Gamma$. The behavior close to the transition energy is similar for both QMEs.
This behavior may be traced back to an energy renormalization which is determined by the imaginary part
of the unilaterally Fourier transformed correlation functions \cite{wubr+05, pela+07},
\begin{equation}\label{eq:MarkovCLTrans}
	C_{\mathbf{a} \mathbf{b}}(\omega) = \int\limits_{0}^{\infty} dt' C_{\mathbf{a} \mathbf{b}}(t') \exp[\pm \imath \omega t']\;.
\end{equation}
In Fig.\ \ref{fig:AppStatCurrU4} the behavior of $C^{(+-)}_{\rm L}$ is exemplarily shown
as a function of the bias voltage. One observes pronounced features
at the transition energies.
In the context of non-Markovian dynamics this can be related to the Lamb-Shift
Hamiltonian, which may lead to deviations from the canonical stationary state \cite{gero+00}.
Explicit time propagation schemes intrinsically account for the renormalization, which
has been discussed in the context of bosonic reservoirs \cite{kl04}.
The non-monotonic behavior observed in the present case indicates the breakdown of the
cumulant expansion with a finite number of terms, which is seen to be valid for couplings 
smaller than the temperature $\Gamma \ll k_{\rm B}T$ \cite{pela+07}.

\subsection{Sudden Switching}
\begin{figure}[tbp!]
  \centering
  \includegraphics[scale=0.39]{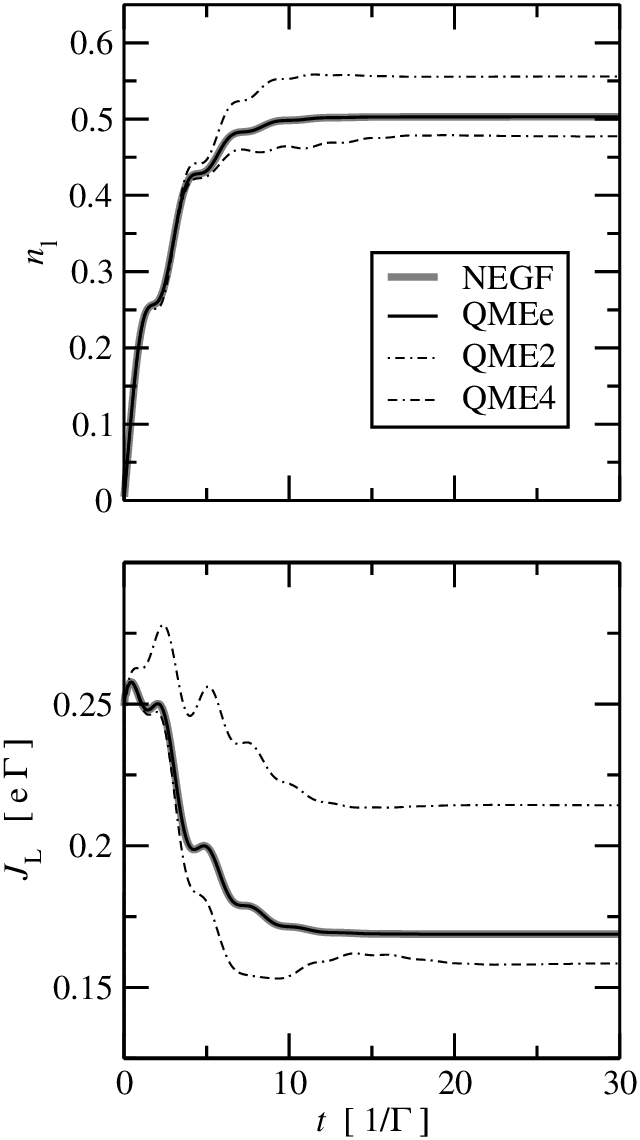}
  \includegraphics[scale=0.39]{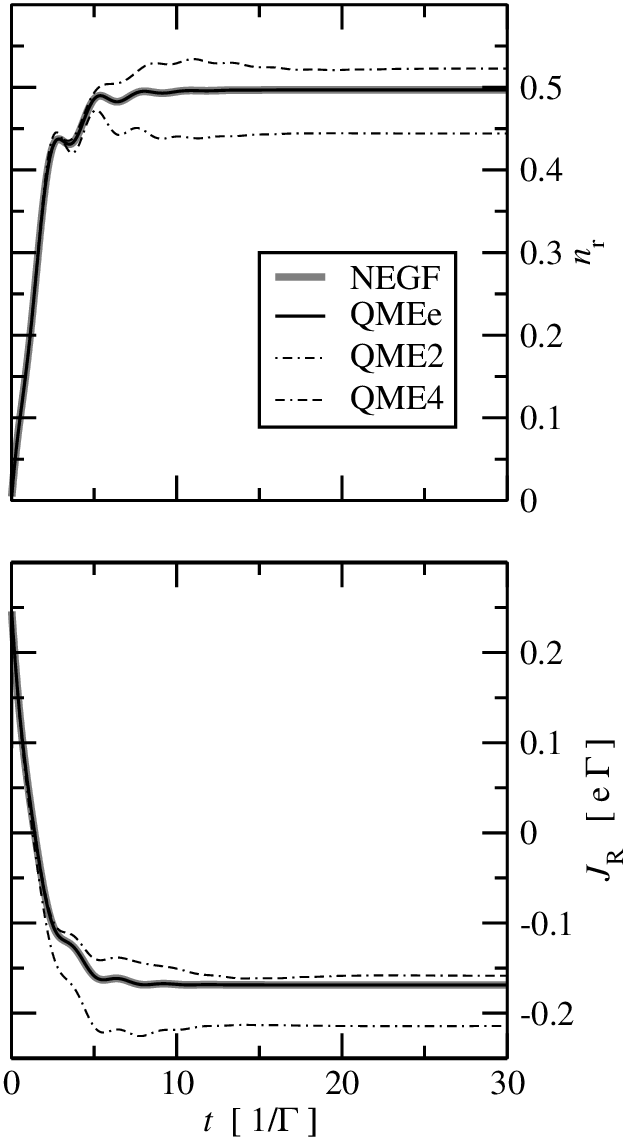}
  \caption{
	  Time-resolved occupations $n_{\ell,\rm r}$ and currents $J_{\rm L,R}$ for $U=0$.
		At $t=0$ the initially empty DQD is suddenly coupled to the reservoirs. 
      Left (right) panel shows occupation of the left (right) dot and 
      the current through the left (right) barrier, respectively. The bias voltage
      is $V_{\rm bias} = 3\Gamma$.
  }\label{fig:AppSuddenVb3}   
\end{figure}

The strength of the formalism given in Sec.\ \ref{sec:wbl} lies in the ability
to study transient responses of the system. Here we consider the time-resolved 
occupation and current after suddenly coupling the initially empty DQD to the reservoirs. 
Figure \ref{fig:AppSuddenVb3} shows the results obtained from the different QMEs for
non-interacting electrons at $V_{\rm bias}=3\Gamma$.
In all cases one observes an initial 
transient response to the sudden coupling and the eventual attainment of a stationary state,
which was discussed in the previous section. Comparing the 
QME results to the exact NEGF calculations one observes that the effective QME indeed yields
the correct dynamics in accordance with Sec.\ \ref{sec:effauxopeom}.
In contrast, the second and fourth order calculations reveal substantial deviations from the exact results. 
In particular, we observe a slower damping of the oscillations, which is expected since the additional
term in the effective QME provides a damping for the \ADO{}. This term is not present in the
second oder case and only partially accounted for in the fourth order case. The
deviations in the long-time limit correspond to the discussion for the stationary
case.

\begin{figure}[b]
  \centering
  \includegraphics[scale=0.7]{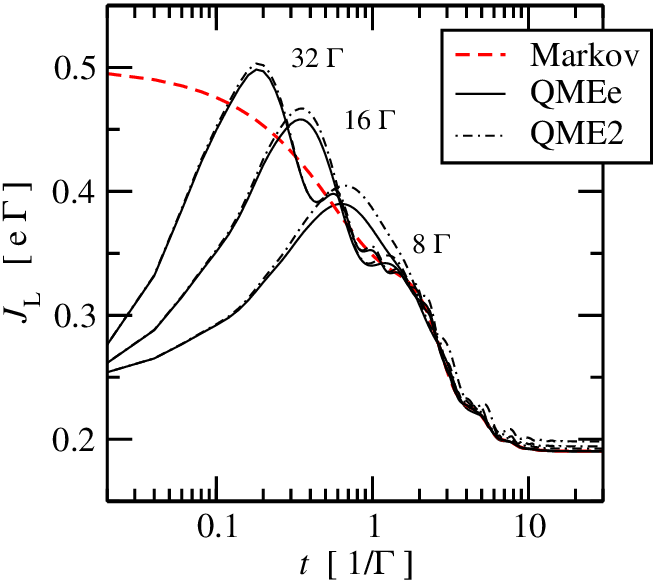}
  \caption{Time-resolved current $J_{\rm L}(t)$ through left barrier for $U=0$
  	and different bias votages $V_{\rm bias} = 8\Gamma, 16\Gamma$ and $32\Gamma$.
	At $t=0$ the initially empty DQD is suddenly coupled to the reservoirs. 
  }\label{fig:AppSuddenVb}   
\end{figure}
Going to larger bias voltages one finds that these deviations vanish and eventually the
Markovian result is obtained. This is illustrated in Fig.\ \ref{fig:AppSuddenVb}. The
difference between the Markovian approximation and the other methods for $t\to 0$ 
are due to the lack of memory in the Markovian approach. This behavior complements the 
discussion of the large bias limit in the previous section.

\section{Summary}
\noindent
We have developed a propagation scheme for the reduced density matrix by means of a time-nonlocal QME. Hereby, we obtained expressions for the memory kernel
up to the fourth order, which has not been discussed before. In order to get
a viable numerical scheme we introduced auxiliary density operators. An expansion
of these operators in terms of auxiliary modes, given by a partial fraction decomposition of the Fermi function \cite{crsa09crsa10}, yields a set of coupled differential equations that can be solved using standard methods.
To investigate the validity of the fourth-order equations we considered the particular case of noninteracting electrons. A comparison with exact results for the electric current
obtained from an NEGF formalism \cite{crsa09a} showed that the fourth-order description is not complete.
Based on this analysis we proposed equations of motion for higher-order
auxiliary density operators, which leads to the correct dynamics of single-particle
observables. Interestingly, these equations only involve quantities with one auxiliary
mode, which is interpreted as a single excitation in the reservoirs. The consistent
treatment of this excitation up to all orders in the tunnel coupling with the contacts leads to a
broadening of the energy levels, which is essential for a complete description of
the dynamics. Further, we compared the proposed effective QME
with the results of a path-integral approach \cite{jizh+08} and found that the QME corresponds to the first level of the hierarchy.

Finally, we showed and discussed the results of the propagation scheme for the specific example of a double quantum dot. Hereby, we verified that the effective QME with a single auxiliary mode is in agreement with NEGF results for noninteracting
electrons. For a finite interaction strength we observed the appearance of unexpected
structures in the current-voltage curve for the second and fourth-order calculations. These
findings indicate the breakdown of the finite-order expansions for strong tunnel coupling.

In total, the combination of auxiliary density operators and the auxiliary mode
expansion yields an efficient method which allows for a 
numerical description of time-resolved electron transport. As we have shown,
the inclusion of higher-order terms is a necessary ingredient for problems involving
strong tunnel couplings or low temperatures.
This opens the possibility to study the response of complex
systems to time-dependent drivings and to investigate realistic schemes for
coherent manipulation of nano-devices.

\appendix
\begin{widetext}
\section{Fourth Order Partial Cumulant} \label{sec:PartCum}
\noindent
The partial cumulants may be defined in terms of a recursion relation \cite{te74,shar80},
\begin{equation}
  \mean{\SO{L}(t) \ldots \SO{L}(\tau_n) \ldots \SO{L}(\tau)}_{\rm pc} 
  ={\sum} ' (-1)^{(g-1)} \mean{\SO{L}(t) \ldots} \mean{\SO{L}(\tau_n) \ldots} \mean{ \ldots \SO{L}(\tau)} \;,
\end{equation}
where the sum is over all partitions and $g$ is the number of partitions in the term. The partitions are such
that the time-ordering $t\ge \ldots \ge  \tau_n \ge \ldots \ge \tau$ is preserved.
Taking $\mean{\SO{L}(t)} = 0$ the fourth order partial cumulant is
\begin{equation}\label{eq:4thpartcum}
  \mean{\SO{L}(t)\SO{L}(\tau_1)\SO{L}(\tau_2)\SO{L}(\tau)}_{\rm pc} 
   = \mean{\SO{L}(t)\SO{L}(\tau_1)\SO{L}(\tau_2)\SO{L}(\tau)}
   -\mean{\SO{L}(t)\SO{L}(\tau_1)}\mean{\SO{L}(\tau_2)\SO{L}(\tau)} \;,
\end{equation}
where the two-point correlation operators are given by \eqref{eq:2ndkernel}. The four-point correlation operators
can be expressed in system and bath operators, by expanding the commutators and taking the partial trace,
\begin{equation}
  \mean{\SO{L}(t)\SO{L}(\tau_1)\SO{L}(\tau_2)\SO{L}(\tau)} \bdot
  = \COMM{S(t)}{S(\tau_1) \SO{F} \bdot - \SO{\tilde{F}} \bdot S(\tau_1)}\;,
\end{equation}
with
\begin{subequations}\label{eq:partcumF}
\begin{align}
  \SO{F} \bdot  = {}
        & \mean{B(t)      B(\tau_1) B(\tau_2) B(\tau) }   \;S(\tau_2) S(\tau) \bdot
        -\mean{B(\tau)   B(t)      B(\tau_1) B(\tau_2) } \;S(\tau_2) \bdot S(\tau) \notag \\
        &-\mean{B(\tau_2) B(t)      B(\tau_1) B(\tau) }   \;S(\tau) \bdot S(\tau_2) 
        +\mean{B(\tau)   B(\tau_2) B(t)      B(\tau_1) } \;\bdot S(\tau) S(\tau_2) \\
  \SO{\tilde{F}} \bdot  = {}
        &\mean{B(\tau_1) B(t)      B(\tau_2) B(\tau) }   \;S(\tau_2) S(\tau) \bdot 
        -\mean{B(\tau)   B(\tau_1) B(t)      B(\tau_2) } \;S(\tau_2) \bdot S(\tau) \notag \\
        &-\mean{B(\tau_2) B(\tau_1) B(t)      B(\tau) }   \;S(\tau) \bdot S(\tau_2) 
        +\mean{B(\tau)   B(\tau_2) B(\tau_1) B(t)      } \;\bdot S(\tau) S(\tau_2) 
\end{align}
\end{subequations}
For self-adjoint reservoir operators $B$ in the expression above one finds $\SO{\tilde{F}} = \SO{F}^\dagger$ (see also
\cite{jaca+02}, where a derivation was given for bosonic operators).

For the case of non-interacting electrons in the reservoir with an initial thermal state, we can further simplify the
four-point correlation functions using the fermion version of Wick's theorem \cite[appendix\,3]{givi05},
\begin{equation}\label{eq:WickFermions}
  \mean{B(t) B(\tau_1) B(\tau_2) B(\tau) }
  = \mean{B(t) B(\tau_1)} \mean{B(\tau_2) B(\tau) }
  - \mean{B(t) B(\tau_2)} \mean{B(\tau_1) B(\tau) }
  + \mean{B(t) B(\tau)} \mean{B(\tau_1) B(\tau_2) } \;.
\end{equation}
Using the correlation super-operators \eqref{eq:corrsupop} and Eqs.~\eqref{eq:partcumF} and \eqref{eq:WickFermions}
one gets
\begin{align}\label{eq:4thcumulant}
  \mean{\SO{L}(t)\SO{L}(\tau_1)\SO{L}(\tau_2)\SO{L}(\tau)} \bdot
  = {} &\COMM{S(t)}{ 
    \SO{C}(t,\tau_1) \COMM{S(\tau_2)}{ \SO{C}(\tau_2,\tau) \bdot} }\notag\\
  &+\COMM{S(t)}{
    \ACOMM{S(\tau_1)}{\left( \SO{A}(\tau_1,\tau_2)\SO{C}(t,\tau)
    -\SO{A}(t,\tau_2)\SO{C}(\tau_1,\tau)\right) \bdot}
    }\;.
\end{align}
Moreover, with the definition of the correlation super-operators, \eqref{eq:2ndkernel} and \eqref{eq:corrsupop},
one finds
\begin{equation}\label{eq:4thordered}
  \mean{\SO{L}(t)\SO{L}(\tau_1)}\mean{\SO{L}(\tau_2)\SO{L}(\tau)} \bdot
  = \COMM{S(t)}{ 
    \SO{C}(t,\tau_1) \COMM{S(\tau_2)}{ \SO{C}(\tau_2,\tau) \bdot} } \;.
\end{equation}
Therefore the fourth order partial cumulant is given by \eqref{eq:4thcumulant} without the first line, which
is canceled by the expression above. This result was used in Eq.~\eqref{eq:4thkernel} for the fourth order memory kernel.
\end{widetext}

\section{Non-Equilibrium Green Functions (NEGF)}\label{sec:NEGF}
\noindent
The description of electron transport by means of
non-equilibrium Green functions has been widely used 
\cite{haja07}
since its first formulation \cite{wija+93}.
Here, we summarize the main equations in a form, which differs
slightly from the usual one, in order to facilitate comparison
with the QME approach of Sect.~\ref{sec:InfiniteDensityMatrix}.
For a more detailed derivation see Ref.\ \cite{crsa09a}.

The current $J_\alpha$ through the barrier connecting 
lead $\alpha$ and the system is given by
\begin{equation}\label{eq:GenCurrent3}
  J_\alpha (t)= {2 e\;} \Re \Tr \left\{ \boldgreek{\Pi}_\alpha (t) \right\}\;.
\end{equation}
with current matrices 
\begin{align}
  \boldgreek{\Pi}_\alpha (t) 
  = {} &  \int\limits_{-\infty}^{t} dt_1 \left[ 
      \mathbf{G}^> (t, t_1) \boldgreek{\Sigma}^<_\alpha (t_1,t)  \right.\notag\\
  &\qquad\qquad \left.-\mathbf{G}^< (t, t_1) \boldgreek{\Sigma}^>_\alpha (t_1,t) \right]\;,
  \label{eq:GenCurrent2}	
\end{align}
describing the flow of electrons from the reservoir into the
system and vice versa. 

By means of the Keldysh equation \cite{haja07} for
$\mathbf{G}^>$ and Dyson-type equations \cite{haja07} for
$\mathbf{G}^\mathrm{a}$ and $\mathbf{G}^\mathrm{r}$,
respectively, one obtains as equation of motion for the density
matrix $\boldgreek{\sigma}(t)\equiv -\imath \mathbf{G}^{<}(t,t)$
\begin{align}
  \imath \frac{\partial}{\partial t}\boldgreek{\sigma}(t)
   = {} & \COMM{\mathbf{h} (t)}{\boldgreek{\sigma}(t)}
  \nonumber\\ &
  +\imath \sum\limits_\alpha \left\{ \boldgreek{\Pi}_\alpha (t) 
    + \boldgreek{\Pi}^\dagger_\alpha (t)\right\} \;,
  \label{eq:ROPDMEOM}
\end{align}
which resembles the structure of a QME.

In order to obtain the time evolution of the current matrices
$\boldgreek{\Pi}_\alpha$ one can write the self energies  
\begin{subequations}
  \label{eq:defself}
  \begin{align}
    \boldgreek{\Sigma}_\alpha^{<}(t_1,t) &= 
    +\imath\mathbf{C}^{(+-)}_\alpha(t,t_1)
    \\
    \boldgreek{\Sigma}_\alpha^{>}(t_1,t) &= 
    -\imath\mathbf{C}^{(-+)}_\alpha(t_1,t)
  \end{align}
\end{subequations}
in full analogy to the correlation functions
(\ref{eq:CorrFunAsSum}) as a finite sum
\begin{align}
  \boldgreek{\Sigma}_\alpha^{<,>}(t_1,t) = {} &  
  \pm\frac{\imath}{2}\boldgreek{\Gamma}_\alpha(t_1,t)\delta(t{-}t_1)
  \nonumber\\
  &\mp\imath\sum\limits^{N_\mathrm{F}}_{p=1}
  \mathbf{C}_{\mathbf{a}\mathbf{b},p} (t_1,t)
\label{eq:selfexp}
\end{align}
with $\boldgreek{\Sigma}_{\alpha p}(t_1,t)
=\imath\,\mathbf{C}_{\alpha p}(t_1,t)$.
Using these expansions in the definition of the 
current matrices, cf.\ Eq.~(\ref{eq:GenCurrent3}), one obtains 
\begin{equation}
\boldgreek{\Pi}_{\alpha}(t) = \frac{1}{4} \boldgreek{\Gamma}_\alpha (t, t)
			-\frac{1}{2} \boldgreek{\sigma}(t) \boldgreek{\Gamma}_\alpha (t, t) 
			-\sum\limits_p \boldgreek{\Pi}_{\alpha p}(t) \;,
	\label{eq:AuxCurrMatDef}
\end{equation}
which defines auxiliary current matrices
$\boldgreek{\Pi}_{\alpha p}$. Using Eqs.\ \eqref{eq:CorrFunAsSum} the equations of motion for 
$\boldgreek{\Pi}_{\alpha p}$ read
\begin{align}\label{eq:AuxCurrMatEOM}
  \imath\frac{\partial}{\partial t}\boldgreek{\Pi}_{\alpha p}(t)
  = {} &\boldgreek{\Sigma}_{\alpha p}(t,t)
  \\
  &+\left(\mathbf{h}(t)-\chi^\pm_p\mathbf{1}-\frac{\imath}{2}\boldgreek{\Gamma}(t,t)\right)
  \boldgreek{\Pi}_{\alpha p}(t)\;.
  \nonumber
\end{align}

\section{Markov approximation} \label{sec:Markov}
\noindent
Since there tends to be some confusion about the precise meaning
of the Markov approximation we give in the following the definitions
used in the present work. Starting with the second order QME, cf.\ Eqs.\ 
\eqref{eq:Def2ndAuxOp} and \eqref{eq:EOMTNLQME}, we first substitute
$\RDM(\tau)$ by $\SO{U}^\dagger_{\rm S}(t - \tau) \RDM(t)$. This renders
the QME local in time and the same equation could have been obtained directly
from the Tokuyama-Mori approach \cite{tomo76}. Next we set the initial time
$t_0 = -\infty$ and also set $t - \tau = t'$ in the integral. The resulting
QME reads
\begin{equation}
  	\imath \frac{\partial}{\partial t} \RDM(t) =
		\COMM{H_{\rm S}}{\RDM(t)} 
		-\imath \,\SO{R}\,\RDM(t)
\end{equation}
with the usual Redfield super-operator (e.g.\ see \cite{maku00})
\begin{align}
  \SO{R}\,\bdot = {} &\sum\limits_{\mathbf{a} \mathbf{b}} \int\limits_{0}^{\infty} dt' \;
			\left[S_{\mathbf{a}},  
                          C_{\mathbf{a} \mathbf{b}}(t') 
    S_{\mathbf{b}}(-t') \bdot 
     \right.\notag\\ & \qquad\qquad\left.
    -  \bdot S_{\mathbf{b}}(-t') C_{\mathbf{b} \mathbf{a}}( -t') \right]_{-}
\end{align}
and $S_{\mathbf{b}}(-t') = \SO{U}_{\rm S}(t') S_{\mathbf{b}}$. Further, using the eigenbasis
of $H_{\rm S}$ one has to evaluate the Laplace transform $C_{\mathbf{a} \mathbf{b}}(E_{A B})$ of the
reservoir correlation functions (cf.\ Eq.\ \eqref{eq:MarkovCLTrans}),
which can be done analytically for the WBL. The final step yielding the Markov
approximation consists in neglecting the imaginary part of the integral.


\end{document}